\documentclass[sigconf,screen]{acmart}
\renewcommand\footnotetextcopyrightpermission[1]{} % removes footnote with conference information in first column
\usepackage{multirow}
\usepackage{enumitem}
\usepackage{makecell}
\usepackage{ragged2e}
\usepackage{CJK}
\usepackage{colortbl}
\usepackage{enumitem}
\usepackage{subfigure}
\usepackage{algorithm}
\usepackage{algorithmic}
\usepackage{hyperref}
\usepackage{xcolor}
\setlist[itemize]{label=$\bullet$, leftmargin=22pt, itemsep=0pt, parsep=0pt, topsep=0pt}
\AtBeginDocument{%
  }

\setcopyright{acmlicensed}
\copyrightyear{2018}
\acmYear{2018}
\acmDOI{XXXXXXX.XXXXXXX}

\acmConference[Conference acronym 'XX]{Make sure to enter the correct
  conference title from your rights confirmation emai}{June 03--05,
  2018}{Woodstock, NY}
\acmISBN{978-1-4503-XXXX-X/18/06}

\begin{document}

\title{Leveraging LLMs to Evaluate Usefulness of Document}
%走向更贴近搜索用户满意度的评估:LLM能否judge文档有用性
\author{Xingzhu	Wang}
\email{wangxingzhu2022@ruc.edu.cn}
\affiliation{%
  \institution{Renmin University of China Gaoling School of Artificial Intelligence}
  \country{China}
}
\author{Erhan	Zhang}
\email{erhanzhang@ruc.edu.cn}
\affiliation{%
  \institution{Renmin University of China Gaoling School of Artificial Intelligence}
  \country{China}
}
\author{Yiqun	Chen}
\email{chenyiqun990321@ruc.edu.cn	}
\affiliation{%
  \institution{Renmin University of China Gaoling School of Artificial Intelligence}
  \country{China}
}
\author{Jinghan	Xuan}
\email{2023201941@ruc.edu.cn}
\affiliation{%
  \institution{Renmin University of China School of Statistics}
  \country{China}
}
\author{Yucheng	Hou}
\email{houyucheng@baidu.com}
\affiliation{%
  \institution{Baidu Inc.}
  \country{China}
}
\author{Yitong	Xu}
\email{yx2376@columbia.edu}
\affiliation{%
  \institution{Baidu Inc.}
  \country{China}
}
\author{Ying	Nie}
\email{nieying02@baidu.com}
\affiliation{%
  \institution{Baidu Inc.}
  \country{China}
}
\author{Shuaiqiang	Wang}
\email{shqiang.wang@gmail.com}
\affiliation{%
  \institution{Baidu Inc.}
  \country{China}
}
\author{Dawei	Yin}
\email{yindawei@acm.org}
\affiliation{%
  \institution{Baidu Inc.}
  \country{China}
}
\author{Jiaxin	Mao}
\email{maojiaxin@gmail.com}
\authornote{Jiaxin Mao is the corresponding author.}
\affiliation{%
  \institution{Renmin University of China Gaoling School of Artificial Intelligence}
  \country{China}
}
\begin{abstract}
The conventional Cranfield paradigm struggles to effectively capture user satisfaction due to its weak correlation between relevance and satisfaction, alongside the high costs of relevance annotation in building test collections. To tackle these issues, our research explores the potential of leveraging large language models (LLMs) to generate multilevel usefulness labels for evaluation. We introduce a new user-centric evaluation framework that integrates users’ search context and behavioral data into LLMs. This framework uses a cascading judgment structure designed for multilevel usefulness assessments, drawing inspiration from ordinal regression techniques. Our study demonstrates that when well-guided with context and behavioral information, LLMs can accurately evaluate usefulness, allowing our approach to surpass third-party labeling methods. Furthermore, we conduct ablation studies to investigate the influence of key components within the framework. We also apply the labels produced by our method to predict user satisfaction, with real-world experiments indicating that these labels substantially improve the performance of satisfaction prediction models.
\footnote{Our code can be found at: https://github.com/3233505833/CLUE}
\end{abstract}






\maketitle
\vspace{-10pt}
\section{Introduction}

%estimator&predicter&judge
% 1、评估在信息检索和推荐系统中扮演着至关重要的角色。通过有效的评估方法，开发人员能够量化系统的有效性，从而识别或者创造更高效的搜索技术。现有的评估方法主要包括在线评估和离线的Cranfield评估方法。在线评估通过实时用户反馈来评估系统性能，能够反映用户的真实体验。Cranfield评估是基于预先标注的相关性数据集进行的离线评估方法，它被广泛使用.% 2、随着克兰菲尔德范式的广泛应用，它引起了一些担忧。
Evaluation plays a crucial role in information retrieval. By employing effective evaluation methods, developers can quantitatively measure system performance, and create more efficient search engine technologies. Existing evaluation methods fall primarily into two categories: online evaluations, using real-time user behavior as implicit feedback, and the Cranfield paradigm, relying on test collections annotated with relevance labels. Among them, the Cranfield paradigm has been widely adopted, but its limitations have raised concerns for some time.

%只用克兰菲尔德和相关性忽略了上下文
% 2. 首先，Cranfield方法评估时，检索系统被评估基于相关性的标注。但目前以用户为中心的满意度逐渐成为黄金标准【】，信息检索也被看成一个用户与信息系统交互的动态过程（添加上下文的必要性，是动态的含有用户的行为）。因此静态的单个的文档的内容相关，不意味着该文档给当前的搜索上下文中的用户带来实际的好处，因此也不意味着用户满意。Mao等人的研究表明，基于相关性计算的评价指标与用户满意度之间的一致性并不高（参考2016年的论文），他认为Cranfield评估无法有效捕捉用户的真实满意度。 
% With the widespread adoption of the Cranfield paradigm, several concerns have emerged. 

% 3. 其次，克兰菲尔德范式面临的最大挑战是建立新测试集的高昂成本。具体来说，在现实世界中，网络上会产生大量新内容，需要不断更新和扩展测试集。然而，随着测试集的扩大，包括文档和相关性标签在内的人力成本也会大大增加。
 Firstly, the cost is substantial for constructing a large-scale, high-quality test collection within the Cranfield evaluation framework. Specifically, in the real world, the continuous generation of new web content requires frequent updates and expansion of test collections ~\cite{hejing}. With the expansion of test collections, the associated human costs increase significantly~\cite{abbasiantaeb2024can}. 

\begin{figure*}[h]
  \centering
  \includegraphics[width=0.9\linewidth]{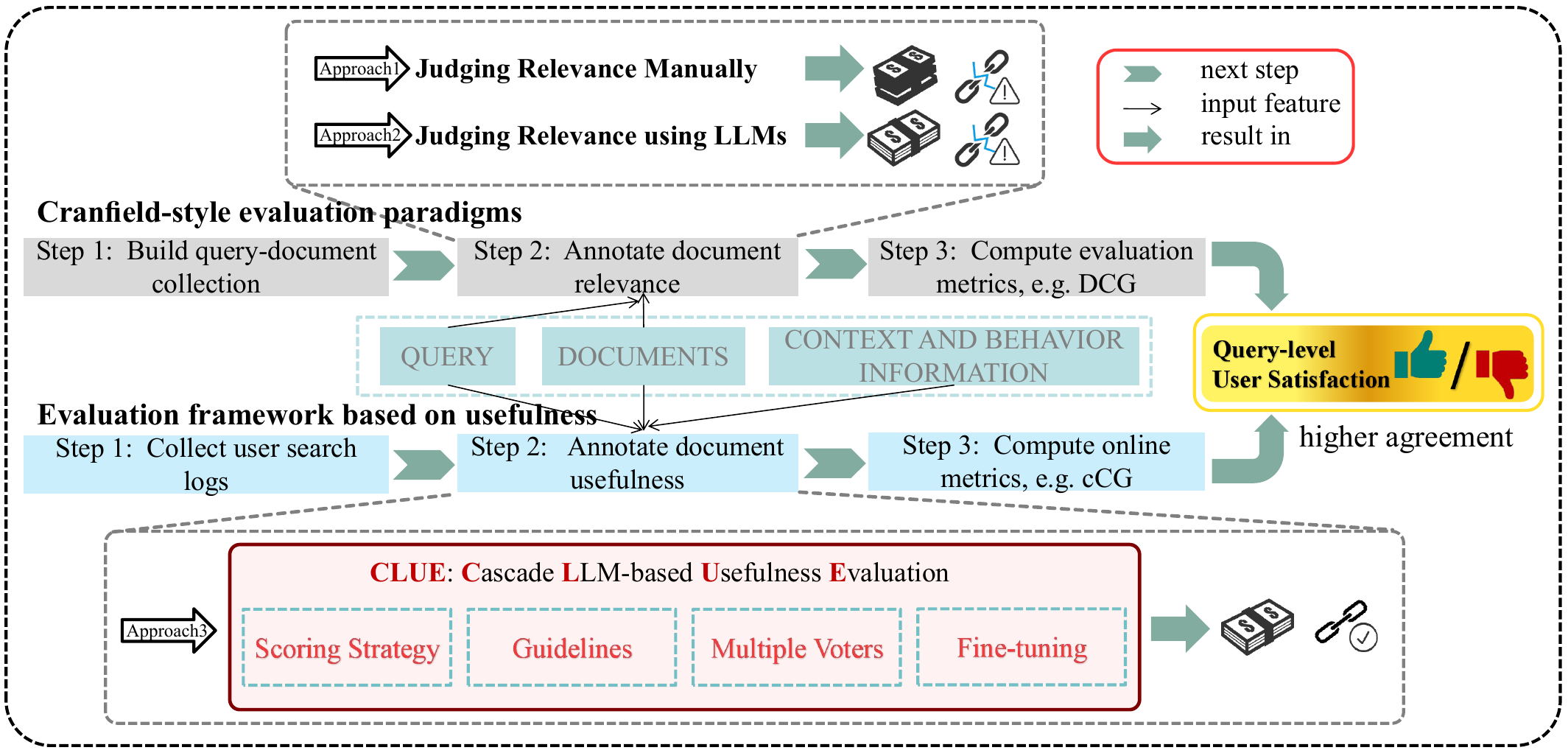}
    \vspace{-3mm}
  \caption{%本图显示了我们的工作背景以及与其他工作的区别。上部子图说明了传统的克兰菲尔德评估和等人倡导的有用性框架，而下部则强调了我们使用 LLMs 评估有用性的独特方法。
Comparison between traditional Cranfield evaluation and the usefulness-based framework proposed by Mao et al. ~\cite{mao2016does}.  %我们在图中比较了三种文档标注的方式。a)使用第三方标注相关性。b)使用大模型标注相关性和c)使用大模型标注有用性。前两种方法由于缺乏用户搜索语境，可能无法设身处地地为用户着想，从而无法确保用户的最终满意度。
  Three approaches in step 2 are: 1) using third-party relevance annotations, 2) using LLMs for relevance annotation, and 3) using LLMs for usefulness annotation. Approach 1 is costly. Approaches 1 and 2, lacking user search context, may not align with user perspectives.
  }
  \label{tab:all}
  \vspace{-17pt}
\end{figure*}

Secondly, the Cranfield evaluation paradigm relies on relevance annotations to assess search engine performance. However, these relevance annotations may not accurately reflect how \textbf{useful} the documents are to users, nor do they capture users' actual satisfaction with the system ~\cite{belkin2015salton, belkin2009model, mao2016does}. Moreover, third-party annotators (also termed silver or bronze annotators) may differ from actual users (also termed gold annotators) in assessing document relevance. This is because third-party relevance annotators focus solely on the relevance matching between the document and topic, and do not have the understanding of the real user’s \textbf{situation}.  This results in a “situational disconnect” between the annotators and users, as illustrated in Figure \ref{tab:all}. Mao et al.~\cite{mao2016does} have shown weak correlations between relevance-based evaluation metrics and user satisfaction, emphasizing their inability  in reflect perceptions from the user’s perspective when evaluating search engines.

To address the first issue, the emergence of LLMs provides a more cost-effective opportunity. LLMs have been widely applied in document relevance assessment~\cite{hou2024large,izacard2022few,khramtsova2024leveraging,faggioli2023perspectives} via pointwise~\cite{zhuang2023beyond}, pairwise~\cite{jiang2023llm,qin2023large}, and listwise~\cite{pradeep2023rankvicuna,sun2023chatgpt} methods. Some studies use LLMs for automatic label generation~\cite{abbasiantaeb2024can,deJesus2024}, while others fill "holes" in evaluation~\cite{abbasiantaeb2024can,macavaney2023one}. Both commercial and open-source LLMs~\cite{touvron2023llama} are explored, with open-source models being less effective but more reproducible and scalable~\cite{macavaney2023one}. 
However, most existing studies focus on relevance annotation with LLMs, with limited attention to user-centric usefulness.

%只用搜索过程和用户行为忽略了文档质量
% 3、针对Cranfield评估的第一个问题，近年来的研究开始关注（加入用户行为）这里需要 强调 基于用户行为做结果级别、查询、任务级别的满意度预测。 jiqun研究了如何用用户的隐式反馈例如用户的点击序列来预测满意度。模型基于用户行为特征和传统机器学习算法的方法在预测满意度很大程度上忽略了结果的内容特征。这忽略了以往的system-centric所注重的文档的质量特征。
To address the second issue, user-centric usefulness from real users can serve as a more reliable and meaningful evaluation metric than third-party relevance labels. However, in a practical Web search setting, asking users to provide explicit feedback on usefulness is impossible. Building upon this challenge, Mao et al.~\cite{mao2016does} have proposed two potential solutions: crowd-sourcing for usefulness label acquisition and supervised machine learning. However, these methods present limitations, as the crowd-sourcing approach is inherently labour-intensive, while the automated machine learning method using bm25 to model the semantic relationship between text content and queries  largely overlooks the content of documents. With the advanced text understanding capabilities of LLMs ~\cite{wei2022chain, wei2022emergent}, which are beyond the reach of small machine learning models they can leverage contextual nuances and semantic relationships to provide more accurate evaluations. 

%以前的fu可以通过第三方标注来实现。也提出了自动化的方法，尤其是毛等人的工作。但首先，他们使用的是qu特征，在很大程度上忽略了的文档的内容。但现在得益于大模型对文档内容的理解能力~\cite{relevance}，我们可以将上下文和文档内容一起输入进去。
%以前使用传统机器学习模型的自动化方法忽略了文档内容。借助已经用于有用性判断的传统机器学习模型所缺乏的高级文本理解能力，LLM可以利用上下文细微差别和语义关系来提供更准确、更人性化的评估。

%但是，在有用性评分这个任务下，我们面临很多额外的挑战：1）我们如何 includes rich behavioral and contextual information when design prompt？2)对于用户来说什么样的文档是有用的？3）我们需要改变打分策略因为用户的往往比相关性列表短，平均长度小于2而不是10，意味着在相关性打分被证明更高性能的pairwise和listwise可能不再适合有用性打分。
% 没有人为有用性的打分发明一个专门的大模型评估方法。没有人说明得到的doc-level的有用性评分将能如何使用来评估搜索引擎？
This naturally leads us to consider employing LLMs for usefulness judgment. But unlike relevance judgment, in the task of usefulness judgment, we face several additional challenges: 1) How can we effectively convey the original users’ search journeys to LLMs when designing prompts? 2) What factors or criteria should LLMs consider when scoring the usefulness of a document? 3) Given that under-evaluation clicked document lists are shorter, averaging less than 2 (Table ~\ref{tab:shuju}), could pairwise and listwise methods, proven outstanding in relevance judgment~\cite{zhang2024large}, still be applicable for usefulness judgment? There is an urgent need to address these issues and design a evaluation method specifically intended for usefulness scoring. 

In response to these challenges, we propose \textbf{C}ascade \textbf{L}LM-based \textbf{U}sefulness \textbf{E}valuation (CLUE).  
 Specifically, 3) inspired by ordinal regression, we treat the usefulness judgment as an ordinal regression problem~\cite{zongshu}. Ordinal regression provides inherent advantages in handling such issues: Unlike nominal classification (which ignores order) and regression (which requires numerical assumptions), ordinal regression is more suitable for predicting categories on an ordinal scale~\cite{35}. Within the broader taxonomy of ordinal regression , we specifically focus on Ordinal Binary Decompositions~\cite{zongshu}. The core idea behind such approaches involves decomposing the ordinal target variable into binary variables, which are then fitted using one or multiple binary classifiers. Regarding decomposition strategies, there exist OneVsNext, OneVsFollowers, OneVsPrevious, and so on~\cite{zongshu}. Regarding binary classifiers, options like logistic regression and SVM are available~\cite{66}. In our method, we adopt the OneVsPrevious approach and implement each binary classifier using the LLM. Our CLUE can be considered an extension of a OneVsPrevious ordinal regression method, where we replace SVMs with LLMs.
Besides the scoring strategy, for 1) and 2), we design prompts that contain rich behavior and context information, incorporating guidelines derived from the newly conducted user study. Furthermore, to enhance performance and robustness, we implement a multi-voter mechanism and integrate the reasoning process. We assess the performance of our approach within the existing usefulness predict methods and demonstrate how the usefulness labels we generate can benefit query-level search evaluation. 
%写我们做了什么。
% 4. 于是我们尝试用LLM来judge usefulness。我们在此过程中综合考虑了文档内容和用户的行为特征。并且我们尝试探索如何更好地利用 LLMjudge 来有用性，以实现更高的一致性，特别是与用户反馈这一黄金标准的对齐，同时提高在不同数据集之间的泛化能力。 具体来说，我们在实验中,首先，We answer a important question:can LLM 可以在检查用户的搜索上下文的辅助下扮演文档有用性法官的角色？接着we interested in  show that the correlatioHow can large models better judge usefulness?
%但是与相关性不同，有用性是一个主观的概念。我们在指导大模型进行有用性打分时，往往需要知道用户对文档的真实想法。这激励着我们做了新的用户实验收集用户在有用性打分时候的主观想法。在这个工作中，用户的有用性得分和思考被收集，这些自然语言描写的决策过程帮助我们理解有用性和相关性的不同，并帮我们更好进行prompt engineer。

%USEFULNESS-BASED和LLM-BASED的满意度评估框架?

%rq1：克兰菲尔德范式的主题相关性可以代替用户体验到的有用性吗？#大模型可以用来标注有用性吗？已经有大模型标注相关性的工作，但是我们觉得主题相关性不能代表用户体验到的有用性，在以用户为中心的评价要求下。所以我们探索了大模型在有用性的能力。大模型直接判断有用性是可以的，是比相关性更贴近用户的感受的。
In summary, we are interested in addressing the following research questions: 

RQ1: Can our LLMs-based method more accurate than the third-party usefulness labeler or the machine learning method?

RQ2: How do different essential components of CLUE affect the performance?%# LLM techniques, and

RQ3: Can we use usefulness judgments from CLUE to improve the accuracy of satisfaction prediction?

Our key contributions are:
%\开始｛逐项列出｝[leftcmargin=*]
% \item我们引入了一个新的用户研究数据集，它捕获了用户在评估文档有用性时基本考虑的内容。我们根据用户定义有用性的特点总结了LLM的有效指南。我们进行了一系列深入的实验，从用户的角度探索LLM在评估文档有用性方面的能力。
% \我们提出了一种基于LLMs的新方法。我们对三个数据集进行了详细的调查
% 研究评分策略、指导方针和各种特征如何影响LLM的有用性判断。
% \我们进一步为红外系统开发了一个自动化的、以用户为中心的评估框架。我们证明，使用CLUE基于LLM的有用性判断的聚合指标可以提高现实世界场景中的查询级评估准确性。
% \结束｛逐项列出｝

\begin{itemize}[leftmargin=*]
    \item We propose a new method named CLUE, which includes components specifically designed for usefulness judgment. We conduct a detailed ablation study to investigate how these components affect performance.
    \item We have collect a new dataset  \footnote{This dataset will be publicly available after the publication of the paper.}. From this, we derive effective guidelines for LLMs to judge usefulness from users’ perspective. Additionally, it can also contribute to future user-centric evaluation efforts.
    \item We further suggest an automated, user-centric evaluation framework using usefulness-based labels. We demonstrate that aggregated metrics from CLUE can enhance the accuracy of query-level evaluation more effectively than relevance labels in the Cranfield paradigm.
\end{itemize}
\vspace{-8pt}
\section{Related Work}
\label{sec:rele}
\vspace{-3pt}
\subsection{Relevance and Usefulness}
\vspace{-2pt}
%文档相关性是克拉菲尔的评测方法中的两个重要组成部分之一。但是随着人们从以系统为中心的评价方式向以用户为中心的评价方式的转变【】，人们逐渐意识到一个搜索引擎的好坏最终取决于用户的体验反馈【】.近年来，克兰菲尔德范式因其衡量标准与用户满意度之间存在差距而受到普遍批评。
Document relevance is a core component of the Cranfield evaluation paradigm. As evaluations shift from system-centric to user-centric approaches ~\cite{wang2024user}, user satisfaction becomes crucial for a search engine's success~\cite{hejing}. The Cranfield paradigm faces criticism for its weak correlation with user satisfaction~\cite{mao2016does}.
%有用性和基于有用性评估的框架是16年的文章中提出来的一个解决方式,是一个包含了用户的情境价值或者说上下文信息的,不同于相关性的全新概念【】。%多个研究探讨了用户实际反馈的有用性和文档相关性的区别【28】【36】【220】【237】。文档有用性包含了用户实际需求和搜索完成效果【20】，以及用户阅读文档后的实际好处【238】，而相关性这个关注文档与查询的匹配度大不相同【47】.
%毛等人积极地提出了新的评估框架，他们表明，基于有用性的框架可以比传统的基于相关性的方法和一些基于用户行为的方法更准确地估计用户满意度。
Using usefulness rather than relevance as evaluation criteria could address this issue~\cite{mao2016does}. Numerous studies have explored the distinction between usefulness and relevance~\cite{kelly2007effects,sanderson2010user,saracevic2022notion}. Document usefulness encompasses the alignment with user needs, task completion outcomes, and the actual benefits derived from the document~\cite{saracevic1988study,zhang2024large,yilmaz2014relevance}, whereas relevance primarily concerns the match between the document and the topic~\cite{huffman2007well}. Mao et al.~\cite{mao2016does} proposed a usefulness-based evaluation framework, showing its superior predictive capability for user satisfaction compared to the traditional relevance-based framework. 
 \vspace{-9pt}

\subsection{Usefulness Judgment using LLMs}
  %使用大型模型来评估有用性是一个新兴领域，目前仍处于探索的早期阶段。
 Using LLMs to assess usefulness of documents is an emerging field that is still in its early stages of exploration.
 %张等人研究如何使用LLMs对候选文档的有用性进行评估。他们通过将prompt中的“相关”替换为“有用”，来选择更有用的文档以构建RAG系统。 我们的研究旨在使用LLMs评估文档的有用性，但我们的目标和方法与张等人不同。我们专注于尽可能还原用户的视角和经历来评估检索系统提供的文档，并在用户的搜索过程中加入更多的上下文因素。在实验部分，我们复现了张等人的三种有用性评分方法作为我们的基准。
Zhang et al.~\cite{zhang2024large} examined using LLMs for judging document usefulness. Their method replaces "relevant" with "useful" in prompts to enhance Retrieval-Augmented Generation (RAG) systems but doesn't focus on user-centric usefulness in IR evaluation. Therefore, they evaluate all documents, not just clicked ones, and their concept of utility differs from the usefulness notion in IR evaluation discussed here \cite{mao2016does}.
 Besides,
%Dewan等人介绍了一种名为TRUE的方法，该方法采用迭代采样和推理来对文档内容和行为进行建模，生成用于评估文档有用性的量规。
Dewan et al.~\cite{dewan2025llm} introduced TRUE, a rubric-based method using iterative sampling and reasoning to model document content and behavior, producing rubrics to evaluate document usefulness.
 %然而，他们仅仅验证了策略在pointwise上的效果，并没有验证在效果更强的其他打分策略[2020年的]上的适用性和提升。 %他们产生了“”等规则。我们这篇，我们直接通过用户实验收集了用户的真实想法，作为对内容的规则。
While they judge usefulness using the pointwise scoring strategy, there remains an opportunity to explore other potentially more effective methods such as listwise, pairwise~\cite{zhang2024large} or other scoring strategy. 
They utilize LLMs to develop rules. In contrast, our work complements this approach by extracting guidelines from the genuine thoughts gathered from users.
\vspace{-8pt}
\subsection{User Satisfaction Prediction}
%用户满意度是信息检索引擎评价的黄金标准，满意度非常重要。用户一旦感到不满意，他们可能会立即转向其他的搜索引擎。搜索引擎需要建立用户满意度模型来提高他们的服务。搜索引擎公司需要建立满意度预测模型来了解、监控用户的满意度状况来提高他们的服务。使用用户搜索日志预测满意度的方法主要是基于用户隐式反馈的满意度预测方法，包括使用隐式反馈来作为用户满意的替代指标和使用机器学习。

%一方面，已经发现用户行为，如搜索任务的持续时间，与用户满意度有相关性。另一方面，红外系统的有效性度量，例如聚合指标 基于相关性和有用性才是用户满意的根本原因。Maskari等人发现，CG和DCG等有效性指标与用户满意度相关。Jiang等人证明了用户行为指标与满意度之间的相关性，以及搜索结果指标与满意度的相关性。Huffman等人{SIGIR-2007-HuffmanH}根据初始搜索查询中前三个结果的文档相关性，发现用户满意度与线性模型之间存在很强的联系。
%mao 发现有用性和query级别满意度之间具有强相关关系。但是，Hassan等人。等人\引用{哈桑2013个性化}曾说，然而， 这些方法需要相关性判断，这可能会受到限制 并且获得成本很高。“，他们用用户的隐式反馈 以预测查询级别满意度。 这样却会丢失文档本身的质量对用户满意度的直接影响。 之前的一些工作都是用用户行为而没有考虑文件内容，当我们利用LLM综合考虑内容和行为得到文档级别的有用标注后，在和query级别的行为一起预测满意度可能会是一个更好的方法。
User satisfaction represents the gold standard for evaluating information retrieval engines ~\cite{huffman2007how,zhang2020models}.  User behavior, like search duration, correlates with satisfaction~\cite{xu2009evaluating}. More importantly, document quality metrics, especially top document quality, direct affect satisfaction~\cite{Moffat2013,Al-Maskari2007,SIGIR-2007-HuffmanH,mao2016does}. In section~\ref{sec:expeRQ3}, we show that LLM-driven document usefulness with CLUE, along with query-level behavior features, effectively predicts satisfaction.
\vspace{-8pt}
\subsection{Ordinal Regression Problem}
%根据顺序标签的处理方式，顺序回归问题通常可以分为三类：忽略顺序信息的标称分类方法[2][52][53]，使用顺序信息构建连续潜在变量的阈值模型[79][80]，以及将顺序问题分解为多个二元子问题的顺序二元分解方法[66][63][71][62][65]。这些方法广泛应用于标签显示自然排名的场景，如用户满意度评级、医疗分级[5]和信用评分[15]，其主要目标是利用类别之间的固有顺序关系来提高预测性能。与名义分类相比，序数回归可以有效减少错误分类引起的序数偏差[35]。与回归方法相比，顺序回归避免了数字标签中固有的测量假设，使其更适合定性排名任务。’
Ordinal regression problems can generally be classified into three categories based on how they handle ordinal labels: nominal classification methods which disregard ordinal information ~\cite{2,52,53}, threshold models which use ordinal information to construct continuous latent variables ~\cite{79,80}, and ordinal binary decomposition methods which decompose the ordinal problem into multiple binary subproblems ~\cite{66,63,71,62,65}. These methods are widely applied in scenarios where labels exhibit a natural level, such as medical grading ~\cite{5} and credit scoring ~\cite{15}, with the core idea of leveraging the inherent ordinal relationships between categories to enhance predictive performance. 
 \vspace{-9pt}
\section{Problem Definition}
\label{sec:problem}
%问题类别：现有的有用性预测工作将之定义为名义分类问题（pointwise的方法直接问属于什么类别），或者是排序问题，以及set问题。现在我们将之定义为有序回归问题。这种方法同样适用于相关性的分级打分。很少有人做有用性的分类。我们的有用性和等人的定义相同，而和等人的定义不同。我们将其他变量也建模进去。

%有用性的特征是文档在多大程度上帮助用户实现他们预期的任务或搜索目标。该指标以用户为中心，结合了文档的实用性，而不仅仅是相关性。与相关性不同，有用性评估以用户为中心，捕捉用户对文档价值和体验的感知，而相关性则关注信息的客观匹配。
Usefulness is defined as the degree to which a document facilitates a users in achieving their intended tasks or search goals. As a user-centric metric, it encompasses the document's practical utility. In contrast to relevance, which focuses on objective information matching, usefulness assessment aims to capture users' perceptions.

%以往相关性判断的定义是在已知QD的情况下判断R，公式描述是R=QD。for each query  $q$, and document content $d$ .%在以往的对user study文档相关性进行标注的惯例中，整个pool就是在一个固定的搜索引擎上，对于这一个q我们所有检索出来的文档。

The relevance judgment problem is classifying documents' relevance above the cutoff, such as @10, into multilevel categories. Formally, let $q$ be a query issued by a user, $\mathcal{D}$ denote the set of 10 documents retrieved by $q$, the relevance judgment problem can be defined as:
\vspace{-6pt}
\begin{equation}
F_R: \{(q, d) \mid d \in \mathcal{D}\} \to \{C_1, C_2, \dots, C_n\}
\vspace{-6pt}
\end{equation}
%与相关性不同，文档在搜索会话中的有用性是高度情境化的{mao2016does}。为了更好地将情景信息传达给大型模型，我们利用所有三种类型的信息来预测有用性：文档内容、用户行为和上下文因素。
Unlike relevance, the usefulness of a document  is highly situational ~\cite{mao2016does}. To better provide situational information to model, previous work typically incorporates user behavior and contextual factors in addition to the content of documents and queries~\cite{mao2017understanding}.
% 只有点击过的文档被考虑，因此我们只判断点击过的文档的usefulness。
Additionally, previous work considers only clicked documents are useful, so usefulness judgments focus solely on clicked documents~\cite{mao2016does}.
Consequently, the usefulness judgment problem is defined as follows: Given a query, clicked documents, and auxiliary information such as user behavior and contextual information, the model should classify the clicked documents into multilevel categories that align  as consistent with \( U_u \) as possible. Formally, 
\vspace{-6pt}
\begin{equation}
\label{eq:classification_function}
F_U: \{(q, d, i) \mid d \in \mathcal{D}_c, i \in I\} \to \{C_1, C_2, \dots, C_n\},
\vspace{-6pt}
\end{equation}
where $\mathcal{D}_c$ represents clicked documents in the query $q$, $I$ denotes other features, including user behavior information and search context information, and $\{C_1, C_2, \dots, C_n\}$ represents ordered usefulness categories, \( U_u \) and the abbreviation for other measures is presented in Table~\ref{tab:key}. %这里来形式化说f和之前的预测的好处，下面说具体的prompt的设计和理念。
\vspace{-8pt}
\begin{table}[h]
\centering
\caption{Key measures in this study. The ground truth is $U_u$, and both *llm and *a are approximate substitutes of it.}
\label{tab:key}
\vspace{-3mm}
\resizebox{\linewidth}{!}{
\begin{tabular}{lp{9cm}}
\toprule
\textbf{Abbr.} & \textbf{Descriptions} \\ \midrule
$U_{\text{u}}$ & Usefulness feedbacks collected from real users. \\ \midrule 
\midrule
$U_{\text{a}}$ & Usefulness annotations made by third-party crowd-sourced labelers. \\ \midrule 
\multirow[c]{3}{*}{$R_{\text{a}}$} & Relevance annotations made by third-party crowd-sourced labelers. \\ 
\cmidrule{2-2}
& Relevance labels commonly used as $R_a$ in the search engine company; hence, we define them as $R^{'}_{\text{a}}$ \\ \midrule
\midrule
$U_{\text{llm}}$ & Usefulness labels  made by CLUE. \\ \midrule
% Usefulness predictions ($U_{\text{ml}}$) & Usefulness predictions made by sota machine learning method. \\ \midrule
$R_{\text{llm}}$ & Relevance labels  made by CLUE where LLM is prompted to assess relevance using the relevance DNA template~\cite{thomas2024large}. It do not include behavior or context features into the prompt, as illustrated in Figure~\ref{tab:prompt}. \\ \bottomrule
\end{tabular}}
\end{table}

\begin{figure}[h]
  \centering
  \includegraphics[width=\columnwidth]{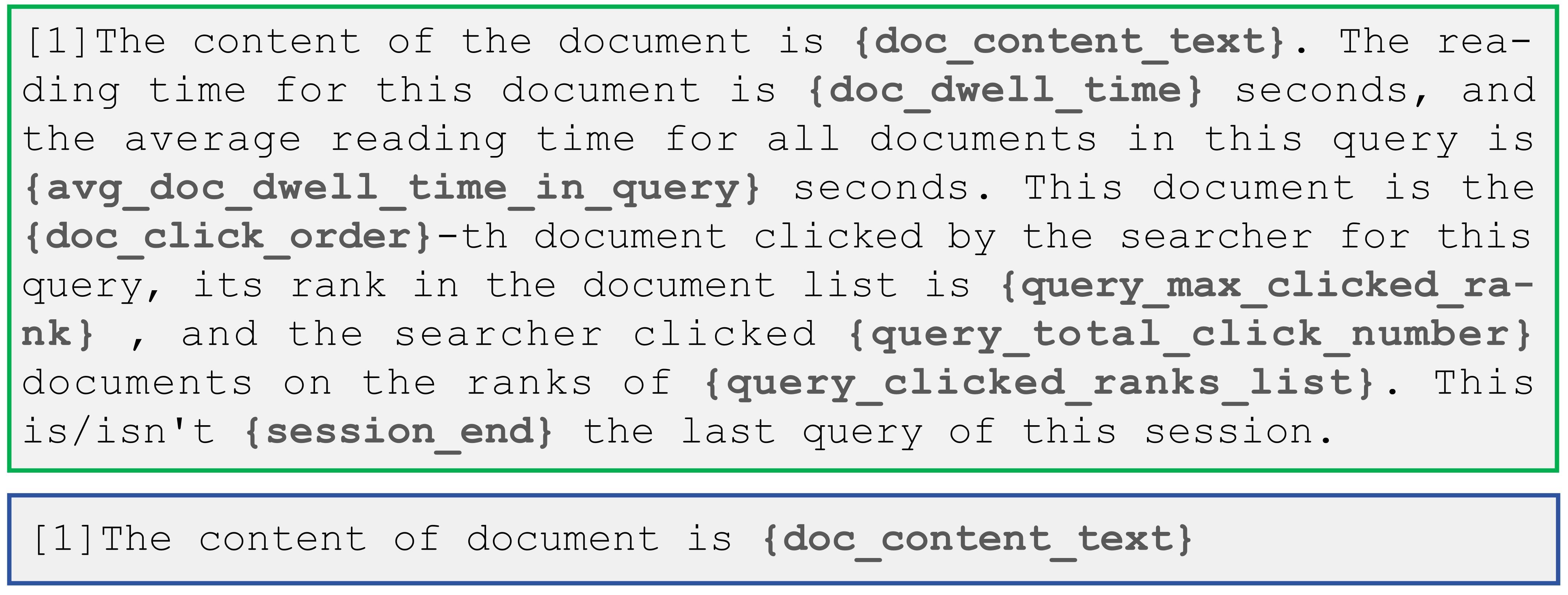}
  \vspace{-20pt}
  \caption{%1）一个search session的样例，用来展示contexutal information包含哪些，我们使用红黄蓝色分别表示doc level，query level和Session l的上下文信息；2）一个Textual description  of Document and Contextual Information,我们使用灰色底色来表示上下文信息所属的类别；3）一个作为对比的不包含contextual information的relevance annotation的prompt
Detail prompt of "[1]document content and other features" in Figure~\ref{tab:prompt_str}.
(1) The green box represents the prompt used for usefulness judgment, where curly braces curly braces indicates the features in Table~\ref{tab:feature_for_llm}.
(2) The blue box represents the prompt for relevance judgment.
}
   \label{tab:prompt}
\vspace{-20pt}
\end{figure}
\begin{figure}[h]
  \centering
  \includegraphics[width=\linewidth]{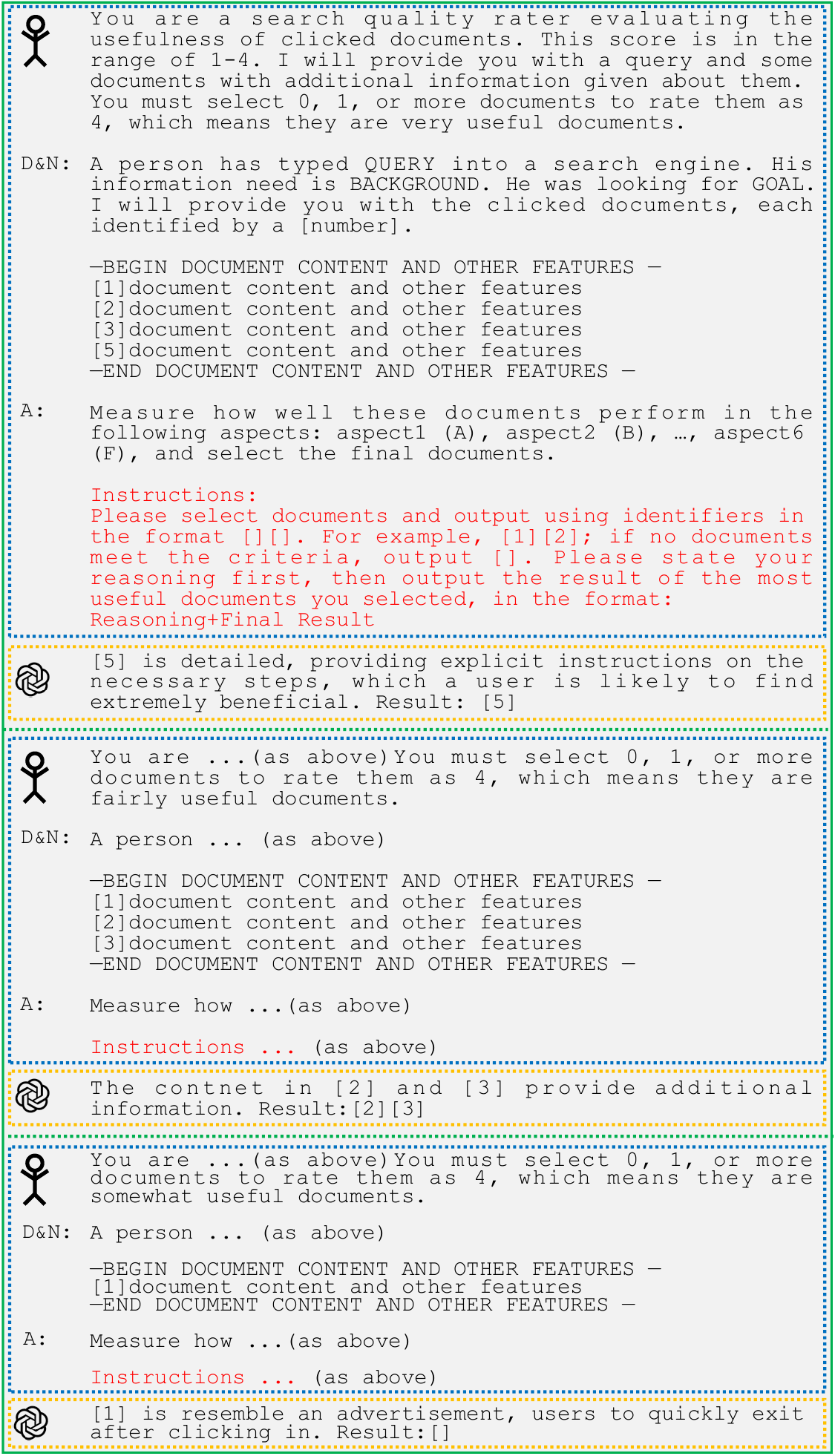}
  \vspace{-20pt}
  \caption{The prompt of CLUE. We use the  DNA prompt template~\cite{thomas2024large}. Capitalized fields like “GOAL” are attributes directly available within the dataset.}
  \label{tab:prompt_str}
\vspace{-15pt}
\end{figure}
\begin{algorithm}[h]
\renewcommand{\algorithmicrequire}{\textbf{Input:}}
\renewcommand{\algorithmicensure}{\textbf{Output:}}
\caption{The Pseudo-code of CLUE with $n$-level Label, $n-1$ Stages and $M$ Voters}
\label{algorithm: iterative_category_assignment}
\begin{algorithmic}[1]
\REQUIRE \textbf{The query $q$, clicked documents set $\{d : d \in D_c\}$, user behavior context information set $\{i : i \in I\}$}\\
\STATE  $X \leftarrow \{x : x \text{ is initialized as } (q, d, i)\}$
\STATE  $Y \leftarrow \{y : y \text{ is initialized as } 0\}$
\FOR{stage $k = n, n-1, \ldots, 2$}
    \FORALL{voters $j$ in $\{1, \ldots, M\}$} 
        \FORALL{\( x \) in \( X \)}
            \IF{$LLM(Prompt_k^j(x))$ is True}
                    \STATE $\text{VoteCount}_k (x) \gets \text{VoteCount}_k(x) + 1$
            \ENDIF
        \ENDFOR
    \ENDFOR
    \FORALL{\( x \) in \( X \)}
        \IF{$\text{VoteCount}_k(x) > \frac{M}{2}$}
            \STATE Assign $y = C_k$
            \STATE \( X \leftarrow X \setminus \{x\} \)
        \ENDIF
    \ENDFOR
\ENDFOR
\FORALL{\( x \) in \( X \)}
    \STATE Assign $y = C_1$
\ENDFOR
\ENSURE \textbf{$n$-level usefulness judgment $\{y : y \in Y\}$ of all clicked document $\{d : d \in D_c\}$}
\end{algorithmic}
\end{algorithm}
% \vspace{-4pt}
\subsection{CLUE}
\label{sec:CLUE}
% 以前的有用性标注方法可以通过第三方注释来实现，也有人提出了自动方法，特别是毛等人的方法。然而，这些方法主要利用查询特征，在很大程度上忽略了文档内容。 利用大型语言模型（LLMs）的高级理解能力（cite{relevance}），我们不再需要特征工程，现在可以将文档内容和上下文信息（包括行为特征）一起输入。
In this section, we introduce CLUE. For simplicity, we use the shorthand $x$ to denote the input tuple $(q,d,i)$ as referenced in Equation~\ref{eq:classification_function} in this \hyperref[sec:CLUE]{section}. %
\vspace{-2pt}
\subsubsection{Overview}
%在有用性评分中，排序信息可用于构建更准确的模型。这确实是一个有序回归问题，但它经常被视为一个标准的名义分类任务，例如在大型模型中使用逐点方法进行文档评分判断时，这可能会导致非最优解。因此，受序数回归的启发，我们提出了这种方法。
In the definition of the usefulness judgment problem, as shown in section~\ref{sec:problem},  the usefulness labels are ordered as $C_1 \prec C_2 \prec \dots \prec C_n$. However, it is often treated as a nominal classification task, such as using GBDT \cite{mao2017understanding}, which potentially leads to suboptimal solutions. The ordering information in the ordered labels can be utilized to construct more accurate models. Therefore, we treat it as an ordinal regression problem \cite{zongshu} in this paper and create the CLUE algorithm.
Algorithm~\ref{algorithm: iterative_category_assignment} shows the Pseudo-code of CLUE. The time complexity of CLUE is $M(n-1)$, since the for loop in lines 5, 11 and 18 essentially executes string matching rather than invoking the LLM for each individual $x$. Next, we present each component of CLUE. 
\vspace{-10pt}
\subsubsection{Scoring strategy}
We start by introducing our motivation. We start by analyzing a method named the Cascade Linear Utility Model, proposed by Wu et al.~\cite{66}, and then present the concept of our CLUE.

\textbf{Cascade Linear Utility Model~\cite{66}}:
In their method, multiple binary classifiers  are constructed, each of them is used to predict each threshold of itself and previous classes. This method determine the category $y$ of $x$ based on the output of the binary classifier $g_{k}(\mathbf{x}; \mathbf{\theta}_{k}$):
\begin{equation}
y 
\begin{cases}
= C_k& \text{if } g_{k}(\mathbf{x}; \mathbf{\theta}_{k}) > 0 \\
 \in \{C_{k - 1}, C_{k - 2}, \ldots, C_1 \}
 & \text{if } g_{k}(\mathbf{x}; \mathbf{\theta}_{k}) \leq 0,
\end{cases}
\end{equation}
where the decision function for each classifier is calculated as:
\begin{equation}
    g_k(\mathbf{x}; \mathbf{\theta}_k)  = w_k^T x + b_k,
\end{equation}
where $w_k$ is the weight vector for the $k$-th model, $b_k$ is the bias, and $\theta_k = \{w_k, b_k\}$ is the parameter set for the $k$-th model.

The prediction process of this method is as follows: for a new sample \( x \), it is sequentially evaluated by a set of binary classifiers, which are arranged in descending order. If \( g_k(\mathbf{x}; \mathbf{\theta}_k) > 0 \), \( x \) is assigned to category \( C_k \); otherwise, it is assigned to a lower category, which can be formally expressed as:
\begin{equation}
\vspace{-5pt}
y = 
\begin{cases}
C_{\arg\max \left\{ k \mid g_k(x; \theta_k) > 0, \ k = n, n-1, \ldots, 2 \right\}}\\
C_1,  \text{if } \forall k, \ g_k(x; \theta_k) \leq 0
\end{cases}
\label{eq:1}
\vspace{2pt}
\end{equation}
\subsubsection*{\textbf{Scoring strategy in CLUE}}
%  In CLUE, we also utilize a chained structure of $n-1$ LLMs to streamline prediction. Given an instance
% $x=(q, d, i)$ and a $n-1 LLM$ , 我们对它确定its predicted label $y$ 建立了以下的算法流程 as follows:
In the CLUE method, we also employ a chained structure consisting of $n-1$ LLMs. The algorithm for CLUE is presented in Algorithm~\ref{algorithm: iterative_category_assignment}. Given an instance $x = (q, d, i)$, and utilizing $g_k(\mathbf{x}; \mathbf{\theta}_k)  =LLM(\text{Prompt}_k(x))$, its predicted $y$ follows:
\begin{equation}
\vspace{-5pt}
y = 
\begin{cases} 
C_k, & \text{if } LLM(\text{Prompt}_k(x)) = \text{True}\\
&\text{\hspace{10pt} for the first} k \text{ from } n, n-1, \ldots, 2 \\
C_1, & \text{otherwise}
\end{cases}
\label{eq:2}
\end{equation}
From Equation ~\ref{eq:1} to Equation ~\ref{eq:2}, and Figure 5 in their work~\cite{66}, it is demonstrated that our CLUE is the generalization of Cascade Linear Utility Model~\cite{66} when LLMs used as binary classifiers. Some may wonder why the process does not start from the lower categories; however, this scoring strategy is evidently symmetric~\cite{66}. The decomposition can also be taken from the other side, where each classifier distinguishes samples belonging to category $C_k$ from those in categories higher than $C_k$.
Specifically, taking the 4-level usefulness judgment task as an example, the estimation is an iterative comparison process. Firstly, we input all $x=(q,d,i)$ in a query at once, as shown in Figure ~\ref{tab:prompt_str}. To determine the category of these $x$, we prompt a LLM, which is seen as $g_k(\mathbf{x}; \mathbf{\theta}_k)$, to select 0,1 or more $x$ that can reach the criterion of $C_4$.  If a $x$ is selected, then the predicted ordinal class of it is $C_4$; For the remaining instances of $x$, we continue to prompt the LLM to determine which $x$ can be selected as  $C_3$. If selected, then the prediction is $C_3$; for the remaining $x$, the process continues. After three stages of selection, the remaining $x$ are all forced to be assigned as $C_1$, as they do not meet the criteria of any of $C_4$, $C_3$, or $C_2$. This way, $4$-level usefulness judgments of all input documents can be obtained. Our prompt is in Figure~\ref{tab:prompt_str}.
\vspace{-6pt}
\subsubsection{Reasoning}
We avoid setting rigid rules based on knowledge from third-party evaluators, such as “dwell time longer than 30 seconds on a content page indicates higher usefulness,” as these assertions are not yet widely accepted axioms and they limit the generalization capability of the method. If inappropriate, they may introduce bias.

So in each prompt, we let LLMs generate a thought first and then output the final selection. The core reason for this prompt design is to not compromise CLUE’ performance on out-of-distribution (OOD) data or scenarios.
\vspace{-6pt}
\subsubsection{Guidelines}

%不像相关性评分一样有详细的打分手册，在以用户为中心的有用性打分中，我们不知道 what users are essentially considering when assessing document usefulness. 这导致我们也无法guides LLMs on how to understand usefulness. Jiqun等人使用逐点输入形式并从附录中导出规则作为指导，这是个好方法。我们的做法是，收集了新的数据并 分析了 users’ thoughts  from think-aloud on document usefulness。具体来说，we tokenize thoughts and 计算了 word frequency. 我们注意到了 six most frequently occurring descriptive adjectives （ \textit{’helpful’}, \textit{’detailed’}, \textit{’relevant’}, \textit{’encyclopedic’}, \textit{’specific’}, and \textit{’comprehensive’}) 并且选择他们作为指导大模型的guideline. 更具体来说，我们 incorporating them into the prompt as aspect in -DNA- for usefulness judgment.
Unlike relevance, there is no guideline  for user-centric usefulness judgment. This leads us to be unclear about what users are essentially considering when assessing document usefulness, making it difficult to guide LLMs. We collect users' thoughts on document usefulness through the think-aloud protocol and analyze them. Specifically, we tokenize their thoughts, calculate word frequency, and rank the frequencies in descending order. Finally, we get the six most frequently occurring descriptive adjectives: helpful, detailed, related, encyclopedic, specific, and comprehensive. We seen them as guidelines and incorporate them into prompt as the \textbf{A}spect in DNA (as shown in Figure~\ref{tab:prompt_str}) for usefulness judgment.
\vspace{-6pt}
\subsubsection{Multiple Voters}
According to the conclusions of Zhang et al. ~\cite{zhang2024large} (in their RQ2), the sequence in which documents are fed into LLMs can affect LLMs’ scoring, potentially leading to some degree of bias. In our method, the order of documents also influence the model’s selection slightly. Therefore, we employ $M$ voters to handle different permutations of sequence at each stage. At stage $k$ the voting count of $M$ voters is given by:
% $$ \text{VoteCount}(x) = \text{Count}\left(\{ j : \text{LLM}(\text{Prompt}^j_n(x)) = \text{True} \text{ for } j = 1, 2, \ldots, M \}\right) $$
\begin{equation}
\vspace{-6pt}
\text{VoteCount}(x) = \sum_{j=1}^{M} \mathbb{I}\left( \text{LLM}(\text{Prompt}^j_k(x)) = \text{True} \right),
\vspace{-1pt}
\end{equation}
where \(\mathbb{I}\) is the indicator function, and \(\text{LLM}(\text{Prompt}^j_k(x))\) is the output of the \(j\)-th prompt with input \(x\). A document is classified as $C_k$ only when it satisfies the following condition:
\vspace{-6pt}
\begin{equation}
y = C_k \quad \iff \quad \text{VoteCount}(x) > \frac{M}{2}
\vspace{-2pt}
\end{equation}
By majority voting considering multiple voters, the variability in each selection can be somewhat diminished, leading to more robust results. 
\vspace{-6pt}
\subsubsection{Fine-tuning}
\label{sec:ft}
%为了优化模型对排名rk的样本与排名较低的样本的判别能力，我们对Llama-3进行了微调。微调开源模型使我们能够为每个子问题构建独立的二进制分类器，而不是依赖于单一的统一模型（例如通过API调用的GPT）进行二进制分类。微调过程利用了每个二进制子问题特有的地面真值（Eq~\ref{}中的y_k）
To optimize the discriminative power for every LLM classifier, we conduct fine-tune. Fine-tuning the open-source model allows us to develop specialized binary classifiers for each subproblem, instead of depending on a singular unified model (such as GPT through API calls). For each stage $k$,  (\( k \in  \{n, n-1, \ldots , 2\}\)), we select the following samples 
\( S_k \) to construct the training set:
\vspace{-4pt}
\begin{equation}
\vspace{-2pt}
    S_k = \{ \{x,y\} \mid y = C_k \} \cup \{ \{x,y\} \mid y < C_k \}.
\end{equation}
\vspace{2pt}  
A training instance consists of an instruction, input, and output. The instruction and input are the same as the string presented in the blue box in Figure~\ref{tab:prompt_str}, while our output is derived from the string shown in the yellow box, but with the reasoning component removed. Each Llama classifier is fine-tuned independently on the corresponding \( S_k \). 
\section{Experimental Settings}
\subsection{Metrics}
\vspace{-3pt}
%我们评估了大型语言模型（LLMs）判断文档有用性的能力，并将评估分为两类： 
%(i) 为了评估用户反馈与所提出的有用性判断模型之间的一致性，我们按等级报告了皮尔逊相关系数（$\rho$）、斯皮尔曼相关系数（Spearman's Rho）和科恩卡帕相关系数（Cohen's Kappa）。我们还可以使用 MAE。
%(ii) 为了评估 LLM 在判断不同类别有用性时的分类能力，我们使用了广泛使用的分类指标，即准确度、宏观精确度（P）、宏观召回率（R）和宏观 F1。
In RQ1 and RQ2, we evaluate various usefulness judgment methods, so we employ two metric types: \textbf{Classification Metrics}: To evaluate classification ability across categories, we use Precision (P), Recall (R), and F1-Score (F1). \textbf{Consistency Metrics}: To assess agreement with ground truth, we report Pearson's $r$ (P-$r$), Spearman's Rho (S-$\rho$), Cohen's Kappa (C-$\kappa$), and MAE. In RQ3, we evaluate satisfaction classification performance using Precision, Recall, and F1-Score.
\vspace{-12pt}
\subsection{The choice of LLMs}
\vspace{-3pt}
%对于商业闭源模型，我们通过可用的 API（包括 gpt-3.5-turbo-16k、gpt-4o-2024-11-20）使用 gpt（openai）。
% 在开源 LLM 中，我们选择了基于 llama 的模型。我们采用的是指令调整版本 Llama-3-8b-chinese-inst。具体来说，我们在以下两种情况下使用 LLAMA：
% (i) 我们直接提示 llama-8b-chinese-inst 生成有用性判断。
% (ii) 我们使用 PEFT 方法 LoRA~\cite{dettmers2023qlora} 对基于 Llama 的模型进行微调，并随后在同一数据集的测试集上对其进行评估。
For the commercial closed-source model, we use GPT (OpenAI) via available APIs, including gpt-3.5-turbo-16k and gpt-4o.
Among open-source LLMs, we choose a instruction-tuned version of llama-based model, Llama-3-8b-chinese-inst. 
\vspace{-12pt}
\subsection{Baseline}
\vspace{-3pt}
\label{sec:baseline}
%我们的baseline总共有两种。在RQ1中我们将现有的文档有用性judgmnet/预测方法作为bl。在RQ2中我们为了说明基于有用性的query-level的评价框架的有效性，我们用了Rllm和Ra^'作为bl。
%For RQ1, to show the effectiveness of our CLUE ,we compare it with existing document usefulness judg/predict methods. 具体来说是the following three types of baselines: 第三方标注的方法，传统机器学习方法，和基于大模型的方法。其中基于大模型的方法我们比较了等人的三种方法。
For RQ1, to show the effectiveness of our CLUE, we compare it with existing document usefulness judgment/prediction methods. Specifically, we use the following three types of baselines: third-party annotation, traditional machine learning, and LLM-based methods. The first is from Mao et al.~\cite{mao2016does} using third-party annotations. The second is their subsequent work~\cite{mao2017understanding} using Gradient Boosting Decision Trees (GBDT) which remains state-of-the-art (SOTA). As the LLM-based usefulness judgment is still in its early stages, our work is one of the pioneering efforts in this area, with limited existing work. Therefore, we compare our method with the following three LLM-based approaches~\cite{zhang2024large}:
%除了pointwise的方法可以直接得到四级分类的分数，其他两种方法pairwise和listwise或直接或间接的yielding the overall ranking of the usefulness of documents.%为了得到四级有用性标签，我们再次指导大模型通过截断排序结果来实现分类。在实验结果分析部分。这种保持了文档之间的相对顺序，因此不会损害相关系数。
\textbf{Pointwise, Pairwise, Listwise}: (i) \textbf{Pointwise}: Input each document tuple  $(q, d, i)$ individually; the output is directly an $n$-level score. (ii) \textbf{Pairwise}: Input each pair as $(q, d_1, i_1)$ and $(q, d_2, i_2)$ within the same query; the output is the one recognized as more useful. Finally, it generate a relative ranking through multiple comparisons. It has the high time complexity of \( O(n^2) \) ~\cite{zhang2024large}. (iii) \textbf{Listwise}: Input the entire set of document tuples $(q, d, i)$ associated with the same query; the output is a ranked list sorted from highest to lowest. In addition to the pointwise, which directly produces $n$-level classification scores, the other two methods—pairwise and listwise—either directly or indirectly provide a ranking like [3]>[2]>[4]>[5]>[1]>[6]. To obtain $n$-level formatted predictions for evaluation, we further instruct the LLMs to segment the rankings into $n$ levels to get a multilevel classification. This post-processing doesn't affect performance metrics as segmentation preserves original ranking order.
For RQ2, to show the usefulness-based evaluation framework's effectiveness, we consider GBDT models additionally inputting $R_{llm}$ or $R_\text{a}/R^{'}_{\text{a}}$ as baselines.
\vspace{-12pt}
\subsection{Dataset}
\vspace{-3pt}
\label{sec:data}
We conducted extensive experiments on four datasets, including two public datasets, SIGIR16 and KDD19, the soon-to-be-released UUST dataset, and a real-world dataset, SearchLog24Q3.
%In RQ1 and RQ2, 我们按照不同的task划分训练集和测试集(5:5)，为了确保在测试时模型没有见过相同的query和document，这一设置对于大模型和机器学习模型都是相同的。统计Data见表2。其中KDD19 和 SearchLog24Q3 used only for testing.In RQ3 ，我们使用GBDT预测满意度的时候,我们对KDD19 和 SearchLog24Q3按照8:2划分并进行了五折交叉验证。 
The data statistics are detailed in Table~\ref{tab:shuju}. For the SIGIR16 and UUST datasets, we split the training and testing datasets by grouping documents based on tasks, ensuring no document or query appears in both sets. The split ratio for the training and testing datasets is 5:5, with 10\% of the training data further partitioned as a validation set. The KDD19 and SearchLog24Q3 datasets are used only for testing. This setup is consistent for both LLMs, traditional machine learning models and third-party annotation methods. In RQ3, when predicting query-level satisfaction using GBDT, we divide queries in KDD19 and SearchLog24Q3 datasets in an 8:2 ratio and performe five-fold cross-validation.
%我们提取这四个数据集共有的特征（如表2所示)，这也是在用户行为数据集中一般都能提取到的特征，也是公司中能轻易收集到的特征，这也保证了我们方法的泛化性。%由于反爬机制，对于三个数据集中的文本内容我们都是通过截屏并且OCR后得到的。对于公司数据集，我们可以轻易拿到文本内容。%这里需要问问要不要加入。这里还是不加入了要不...反正表里写的提取特征已经够清楚了。
\begin{table*}[t]
\centering
\caption{Dataset Statistics. "Query", "Document", and "Clicked documents" are abbreviated as "Qry", "Doc", and "Ck Doc". Moreover, "a laboratory study" and "an anonymous search engine company" are abbreviated as "lab" and "company"}
\vspace{-10pt}
\label{tab:shuju}
\resizebox{\textwidth}{!}{
\begin{tabular}{lccccccccccc} 
\toprule
 & \textbf{Data Source} & \textbf{\# Tasks} & \textbf{\# Users} & \textbf{\#Qry} & \textbf{\# Doc} & \textbf{\#Ck Doc} & \textbf{\#Ck Doc/\#Qry}& \textbf{Difficulty }&\textbf{\#Train Doc}&\textbf{\#Valid Doc}&\textbf{\#Test Doc}\\
\midrule 
\textbf{SIGIR16} & lab & 9 & 25 & 935 & 3K & 1,512  &1.6&Normal & 775& 86 &651 \\
\textbf{KDD19} & lab\textsuperscript & 9 & 40 & 1,548 & 15K & 3,274 &2.1&Complex & - & - &1,826\\
\textbf{UUST} & lab& 10 & 31 & 732 & 7K & 1,425 & 1.9&Normal & 607& 67&751\\
\textbf{SearchLog24Q3} & company & - & 7,572 & 9,547 & 95K & 14,420 &1.5& Daily & -& -& 14,420\\
\bottomrule
\end{tabular}}
\vspace{-16pt}
\end{table*}
\begin{table}[h]
 \centering
 \caption{Scales of key measures.}
 \vspace{-10pt}
 \resizebox{\linewidth}{!}{
  \begin{tabular}{lllll}
   \hline 
   \textbf{Variable}&\textbf{SIGIR16} & \textbf{KDD19} & \textbf{UUST} & \textbf{SearchLog24Q3} \\
   \hline 
   Relevance ($R_a$) & 4-level & 4-level & Not collected & Continuous ([0,1]) \\
   Usefulness ($U_a$) & 4-level & Not collected & Not collected & Not collected \\
   Usefulness ($U_u$) & 4-level & 4-level & 5-level & Not collected \\
   Satisfaction ($QSAT$) & 5-level & 5-level & 5-level & 4-level \\
    \hline
  \end{tabular}}
 \label{tab:variable_defs}
 \vspace{-12pt}
\end{table}
\vspace{-7pt}
\subsubsection{KDD19 Dataset}
%我们使用的是 KDD19 数据集，这是由等人在 2019 年收集的一个公共搜索行为数据集。KDD19 数据集包含 1K 多个查询和 10,000 多份文档。该数据集是通过合作研究收集的，其中 30 名参与者需要完成 9 项复杂的搜索任务。 我们采用了与 Zhang 等人引用[]相同的数据过滤规则，由于大量无效 URL 无法复制用户最初浏览的内容，我们删除了四项任务。此外，我们还排除了记录不完整或缺少某些点击数据的会话。我们在所有评估中都采用了相同的会话。
The KDD19 dataset is a public search behavior dataset from a lab study, detailed in ~\cite{DBLP}. We remove four tasks following the same principles and for the same reasons as outlined in Zhang et al.'s work on user simulation~\cite{zhang2024usimagentlargelanguagemodels}
: We can’t obtain the original document content in these four tasks and we cannot predict usefulness, since these four tasks involve past news searches, with nearly all the URLs now invalid (404). 

\vspace{-7pt}
\subsubsection{SIGIR16 Dataset}
%SIGIR16数据集是通过实验室研究收集的公共搜索行为数据集。在这项研究中，25名参与者完成了9项搜索任务，产生了225个搜索会话日志。对于每项任务，参与者从初始查询的搜索引擎结果页面（SERP）开始，可以自由点击结果或提交新的查询。当他们觉得任务完成或找不到进一步的有用信息时，他们可以结束会议。完成任务后，参与者回顾了他们的搜索过程，并提供了明确的反馈，包括4级有用性评估。
The SIGIR16 dataset is another public search behavior dataset from a lab study.
 %我们使用这个数据集的主要原因有两点：1.里面有U_a字段，我们根据这个数据集可以比较U——llm和U——a的性能。2.这是是有用性预测机器学习方法提出的论文中用的数据集。%问问有没有介绍要不要把网址放进去。
There are two primary reasons for using this dataset: (i) It contains the $U_a$ field, allowing us to compare the performance of $U_{llm}$ and $U_a$ based on this dataset. 
(ii) It is the dataset for usefulness prediction research with machine learning~\cite{mao2016does,mao2017understanding}, enabling us to replicate the machine learning method and compare it with the LLM-based method.
\vspace{-8pt}
\subsubsection{UUST Dataset}
 This dataset is specifically collected for this study. Key differences from previous datasets include: (i) storing complete document text content that was originally viewed by users via screenshots and OCR to address the issue of numerous invalid URLs in existing datasets, and (ii) capturing users’ real-time thoughts during their usefulness rating to gain insights into how to assess usefulness in this and future work. Detailed information is in Appendix~\ref{tab:Dataset Collection}.
\vspace{-7pt}
\subsubsection{SearchLog24Q3}
%%%%% jiang2015understanding
%%%%%
%我们用了2024年7-10月份总共三个月的交互日志。任何个人隐私的信息在分析之前从日志中删除。这些日志条目主要包括每个查询的唯一标识符、用户行为的时间戳，和访问的网页的URL。%接下来query级别的满意度被专家标注。尽管不是来自第一方的标注，但已有的工作证明也能得到可靠的结论。In the real scenario, 从用户那里获得高质量的满意度标注很难除了在实验室中，因此百度就是一直用的专家标注的满意度数据来建立满意度模型的。因此我们也使用了这一标注。%尽管不是来自第一方的标注，但许多先前的研究使用类似的数据也得到了可靠的结论。%这些搜索任务的就是用户现实中的的分布，在这个数据集上能验证大模型标注的有用性在大规模现实情况下的应用价值。
This data is industrial search data collected during the third quarter (July to October) of 2024. Log entries include query IDs, user behavior timestamps, and visited URLs, which are later used to extract user behavior features and document content, respectively. We annotate query-level satisfaction of the search queries by experts (silver annotators). For each query, we present the annotators with the complete user search process to obtain reliable annotations.
Similar to previous studies \cite{hassan2012semi, hassan2010beyond, wang2014modeling, jiang2015understanding} that derive reliable conclusions from expert annotations, we also utilize these annotations as the query-level satisfaction ground truth. This large-scale and real-world dataset enables us to validate the application value of usefulness judgments from LLMs.
% \subsubsection{Ethics.}  在数据收集过程中，我们对百度搜索日志数据进行了匿名处理，以维护用户隐私。查询使用唯一的查询 ID 进行聚合，从而防止研究人员识别单个用户的查询。所有数据收集均在用户知情同意的情况下进行，不会对用户造成任何身体或心理伤害。
\vspace{-6pt}
\subsubsection{Ethics}  During our data collection process, real industrial search log data were anonymized to maintain user privacy. All data collection was conducted with the informed consent of the users, and it did not cause any physical or psychological harm to them.

%待加4.5和4.6，这里要对应问题定义4.1里介绍的内容，说我们要使用LLM来实现f_U这个函数，然后关键在于要清晰的描述具体的过程，并且要和之前LLM for relevance annotation以及传统基于用户行为数据来做usefulness预测的差别，同时回应一下introduction里说的LLM-based usefulness prediction可以解决现有工作的不足

\vspace{-8pt}
\section{Experiments \& Results}
\label{sec:expe}
%在本节中，我们将详细描述为回答每个研究问题而设计的实验，然后介绍实验结果。我们将测试第 4 节中介绍的方法。 As described in Section ~\ref{sec:data}, we use 5 tasks from the KDD19 dataset and 5 tasks from the UUST dataset as the test sets, with the test sets for relevance and usefulness judgments remaining the same.
\subsection{RQ1: Exploring the performance of LLM-based CLUE}
\subsubsection{Experimental setup}
When implementing our CLUE method, we set the LLMs' temperature $t=0$ to reduce uncertainty and set $M=5$. Additionally, we reproduce two existing methods for predicting usefulness (as shown in section~\ref{sec:baseline}) on SIGIR16 dataset , because it is the only dataset that includes $U_a$. Details are in Appendix~\ref{tab:task_details}. $GBDT_{ood}$ refers to the model using the training and test sets defined in this paper as shown in Table~\ref{tab:shuju}. Additionally, we test $GBDT_{iid}$, which involves randomly splitting the documents like their work~\cite{mao2017understanding} into a 5:5 ratio for training and testing.

\begin{table}[ht]
    \centering
    \caption{Extracted features for LLM prompt.}
    \vspace{-10pt}
    \resizebox{\columnwidth}{!}{
    \begin{tabular}{@{}lll@{}}
        \toprule
        \textbf{Content Factors} & \textbf{Context Factors} & \textbf{Behavior Factors} \\ \midrule
        doc\_content\_text & query\_total\_click\_number & doc\_click\_order \\
        query\_string\_text & query\_clicked\_ranks\_list & doc\_dwell\_time \\
        task\_description\_text & query\_max\_clicked\_rank & session\_end \\
        & avg\_doc\_dwell\_time\_in\_query & \\ \bottomrule
    \end{tabular}}
    \label{tab:feature_for_llm}
    \vspace{-12pt}
\end{table}

\begin{table}[h]
\centering 
\caption{Overall performance and comparison between third-party usefulness annotations and the SOTA machine learning method. CLUE-3.5/4o is CLUE based on  GPT-3.5/4o }
\vspace{-10pt}
\label{tab:gpt1}
\resizebox{\columnwidth}{!}{
\begin{tabular}{lccccccc}
\hline
  & Precision & Recall & F1-Score  & P-$r$ & S-$\rho$ & C-$\kappa$ & MAE\\ 
\hline
\textbf{SIGIR16} & & & & & & & \\
$R_a$ & 0.4045 & 0.3018 & 0.3206 & 0.3198 & 0.3434 & 0.1050 & 1.0239\\
$U_a$ &  0.3864 & 0.3780 & 0.3755 & 0.3536 & 0.3517 & 0.1682 & 0.9210\\
$GBDT_{iid}$~\cite{mao2017understanding} &  0.4026& 0.4422 & 0.4396 & 0.4298 & 0.4194 & 0.2392& 0.8613\\
$GBDT_{ood}$~\cite{mao2017understanding}& 0.4003 & 0.4003 & 0.3728 & 0.3157 & 0.3053 & 0.1642&1.0050\\
CLUE-3.5 & 0.3419 & 0.3988 & 0.3556 & 0.3739 & 0.3711 & 0.1611 & 0.9924 \\
CLUE-4o   & 0.3781   & 0.4135 & 0.3813  & 0.3830     & 0.3816     & 0.1845    & 0.9841 \\
\hline
\textbf{KDD19} & & & & & & & \\
CLUE-3.5  & 0.3837 & 0.3805 & 0.3802 & 0.3159 & 0.3177 & 0.1511 & 1.0071 \\
CLUE-4o   & 0.3871& 0.3794&0.3668 &0.3457 & 0.3427&  0.1552& 1.0489\\
\hline
\textbf{UUST} & & & & & & & \\
CLUE-3.5 & 0.4395 & 0.4252 & 0.4233 & 0.3751 & 0.3823 & 0.1911 & 1.1776 \\
CLUE-4o & 0.4419 & 0.4567 & 0.4448 & 0.4275 & 0.4245 & 0.2273 & 1.1603 \\
\hline
\end{tabular}
}
\vspace{-12pt}
\end{table}
\vspace{-7pt}
\subsubsection*{\textbf{Overall Performances}:}
The overview of the results is shown in Table ~\ref{tab:gpt1}.
%首先我们展示了，在克兰菲尔德范式中的第三方标注的相关性和第一方报告的有用性，差距是最大的。因为仅仅代表了主题相关性，而且第三方标注无法带入用户的鞋子。另外，先前的工作成功的使用第三方的有用性标注作为Uu的替代，我们的结果表示，指导良好的LLM可以超过第三方的有用性标注。
Our findings illuminate:
(i) $R_a$ shows the weakest performance, as third-party relevance annotation only reflect topic relevance and lack user perspective.
(ii) CLUE achieves moderate positive correlations across all datasets. GPT-4 outperforms GPT-3 in most cases.
(iii) CLUE can surpass third-party annotations when use GPT-4o (P-$r$ = 0.38 > 0.35 in $U_a$).
%这说明当我们将用户的信息获取过程通过语言对用户的行为和上下文factor进行还原时，大模型能放在用户的鞋子里对user-perceived usefulness进行judge。
This indicates that by reconstructing the user’s information-seeking process through behavioral and contextual factors in language, LLMs can effectively evaluate user-perceived usefulness.
(iiii) The machine learning method typically lose performance from iid to ood setting, whereas zero-shot LLM-based method is not affected by this issue. Additionally, CLUE's performance is comparable to the machine learning method in task ood setting in this paper. This suggests the potential utility of directly using LLM-based methods for usefulness prediction, saving  the costs of collecting training data.
% First,  the relevance of third-party annotations  is 最差的。  
% 我们发现CLUEachieved moderate positive correlations 在所有的数据集上,and  we find that GPT-4 performs better than GPT-3 in most scenarios, but not always.
% Moreover,  our results indicate that well-guided LLMs can surpass the third-party annotations(P-r=0.38当使用GPT4时>0.35 in $U_a$),. 
\vspace{-10pt}
\subsection{RQ2: Ablation study}
\label{rq2}
% rq2: How do different LLM techniques, and essential components of our LLM-based method affect the performance?
%在这里，我们重点回答我们的rq2。为了探索我们基于LLM的方法的基本组成部分的影响，我们进行了一项消融研究。
Here, to explore the effect of the essential components of CLUE, we carry out an ablation study.
%我们使用grade-u模型作为完整模型，它包含了为实现最佳性能而设计的所有功能和组件。为了了解每个组成部分的贡献，我们探索了以下变体，包括LLM和提示策略的选择，以及LLM是否针对有用性预测任务进行了微调：为了减少推理中的不确定性和不稳定性，我们设置温度$t=0$。并且为了纯粹的比较打分策略，在这一节中，我们保证输入的特征不变。
\vspace{-7pt}
\subsubsection{Common Experimental Setup}
In this section, we use our CLUE model as the full model, which incorporates all the features and components designed for optimal performance. To reduce the uncertainty, we set the LLMs' temperature $t=0$. For a fair and pure component comparison, all variants are using the same features (as shown in Figure~\ref{tab:feature_for_llm}) as those in CLUE.
\vspace{-7pt}
\subsubsection{\textbf{Effects of Multi-voter}}
%我们进行了一项消融研究，以验证在每个阶段多选民的必要性。
We conduct an ablation study to verify the necessity of multi-voter mechanism at each stage. The variant with only single voter per stage  is called CLUE-s.
\begin{table}[ht]
\centering
\caption{Performance comparison of CLUE-s with variants using the baseline scoring strategy using GPT-4o.}
\vspace{-10pt}
\label{tab:strategyvariants}
\resizebox{\columnwidth}{!}{
\begin{tabular}{p{2cm}ccccccc}
\toprule
\textbf{Dataset} & \textbf{Metric} & \textbf{Pointwise} & \textbf{Pairwise} & \textbf{Listwise} & \textbf{CLUE-s} \\ 
\midrule
\multirow{7}{*}{\textbf{SIGIR16}} 
& Precision & \textbf{0.4598} & \underline{0.3749} & 0.3715 & 0.3604 \\
& Recall & 0.2869 & \underline{0.3250} & 0.3131 & \textbf{0.3989} \\
& F1-Score & 0.3073 & \underline{0.3257} & 0.3171 & \textbf{0.3663} \\
& P-$r$  & \underline{0.2727} & 0.2701 & 0.2471 & \textbf{0.3650} \\
&S-$\rho$  &\underline{0.2981} & 0.2667 & 0.2444 & \textbf{0.3616} \\
& C-$\kappa$  & \underline{0.1187} & 0.1152 & 0.0999 & \textbf{0.1637} \\
& MAE & 1.087 & \underline{1.012} & 1.038 & \textbf{1.011} \\
\midrule
\multirow{7}{*}{\textbf{KDD19}} 
& Precision & 0.3018 & 0.3420 & \underline{0.3611} & \textbf{0.3658} \\
& Recall & 0.1961 & 0.2531 & \underline{0.3056} & \textbf{0.3608} \\
& F1-Score & 0.2148 & 0.2386 & \underline{0.3188} & \textbf{0.3449} \\
& P-$r$ & 0.1200 & 0.1653 & \underline{0.2251} & \textbf{0.2960} \\
& S-$\rho$  &0.1278 & 0.1641 & \underline{0.2237} & \textbf{0.2944} \\
& C-$\kappa$  & 0.0247 & 0.0445 & \underline{0.0811} & \textbf{0.1316} \\
& MAE & 1.4938 & 1.2209 & \underline{1.0760} & \textbf{1.1148} \\
\midrule
\multirow{7}{*}{\textbf{UUST}} 
& Precision & \underline{0.4054} & 0.3123 & 0.3138 & \textbf{0.4406} \\
& Recall & 0.2177 & 0.2573 & \underline{0.2615} & \textbf{0.4448} \\
& F1-Score & 0.2237 & 0.2613 & \underline{0.2634} & \textbf{0.4365} \\
& P-$r$ & \underline{0.3077} & 0.2974 & 0.2829 & \textbf{0.4159} \\
& S-$\rho$  & \underline{0.3267} & 0.2888 & 0.2797 & \textbf{0.4121} \\
& C-$\kappa$  & 0.0681 & 0.0750 & \underline{0.0805} & \textbf{0.2106} \\
& MAE & 1.4984 & \underline{1.3091} & 1.3189 & \textbf{1.1748} \\
\bottomrule
\end{tabular}
}
\vspace{-16pt}
\end{table}
\vspace{-7pt}
\subsubsection*{\textbf{Results}:} 

%多版本方法优于单版本方法，表明多个评分者可以提高准确性。这些评分器旨在对每个阶段输入列表中文档的位置保持敏感。尽管这增加了算法的时间复杂度，但每个阶段的错误都会显著影响级联过程中的后续阶段。因此，采用这种策略来降低敏感性是值得的。
As shown in Table~\ref{tab:gpt2}, the multi-version approach outperform the single-version method, indicating that multiple voters can enhance accuracy. The main function of having multiple voters at every stage is to reduce sensitivity to the position of documents within inputs. Although this approach increases the algorithmic time complexity, it significantly reduces the errors at each stage. Therefore, employing this strategy is worthwhile.
\begin{table}[h]
\centering
\caption{Ablation study result on the multi-voter component. "CLUE-s" is an abbreviation for the variant with only single voter per stage.}
\vspace{-10pt}
\label{tab:gpt2}
\resizebox{\columnwidth}{!}{
\begin{tabular}{lcccccccc}
\hline
 & Precision & Recall & F1-Score  & P-$r$ & S-$\rho$ & C-$\kappa$ & MAE\\ 
\hline
\textbf{SIGIR16} & & & & & & & \\
CLUE   & \textbf{0.3781}   & \textbf{0.4135} & \textbf{0.3813}  &\textbf{ 0.3830  }   & \textbf{0.3816 }    & \textbf{0.1845}    & \textbf{0.9841} \\
CLUE-s &0.3604 & 0.3989 & 0.3663 & 0.3650 & 0.3616 & 0.1637 & 1.011 \\
\hline
\textbf{KDD19} & & & & & & & \\
CLUE   & \textbf{0.3871}& \textbf{0.3794}&\textbf{0.3668} &\textbf{0.3457} & \textbf{0.3427}&  \textbf{0.1552}& \textbf{1.0489}\\
CLUE-s &0.3658 & 0.3608 & 0.3449 & 0.2960 & 0.2944 & 0.1316 & 1.1148 \\
\hline
\textbf{UUST} & & & & & & & \\
CLUE & \textbf{0.4419} & \textbf{0.4567} & \textbf{0.4448} &\textbf{ 0.4275} & \textbf{0.4245} & \textbf{0.2273} & \textbf{1.1603} \\
CLUE-s &0.4406 & 0.4448 & 0.4365 & 0.4159 & 0.4121 & 0.2106 & 1.1748 \\
\hline
\end{tabular}
}
\vspace{-12pt}
\end{table}
\vspace{-7pt}
\subsubsection{\textbf{Effects of Scoring Strategy}}
%%
%我们对我们的打分与多个现有的有用性打分策略的基线进行了比较分析。结果见表3。为了确保比较的公平性，所有基线均输入和我们一样多的特征信息重新实现。
We compare our scoring strategy with baselines with existing usefulness scoring strategies as shown in section~\ref{sec:baseline}. To ensure fair comparisons and reduce API resource consumption, we use one voter for all methods. 
We establish variants by replacing the scoring strategy component with the three baseline scoring strategies.

%观察到Pairwise和Listwise方法在所有数据集和所有设置上都明显优于Pointwise方法。}这可能是因为，与独立处理每个判断的Pointwise方法不同，Pairwise和Listwise方法利用了LLM识别实例之间偏好的能力，使其能够提供更准确的选择或评级。
\vspace{-7pt}
\subsubsection*{\textbf{Results}:} The results are shown in Table ~\ref{tab:strategyvariants}. We can observe that: (i)scoring strategy of CLUE achieves state-of-the-art performance, significantly outperforming three baseline methods across all datasets. Furthermore, the consistency metrics significantly exceed those of the baseline methods, proving the method’s effectiveness in preserving ordinal relationships. 
% 最后，对比三种方法我们发现，LLMs在KDD19数据集上的性能数值最低。因为（KDD19 from a laboratory study "in which users need to complete some complex search tasks"\footnote{kdd1}），无论是大模型还是人都很难辨别一个文档是否对困难任务有用。另外，我们发现listwise方法在更长的文档列表within a query上会发挥出它的性能，
%KDD19的困难属性导致它的query平均点击长度更长，这会使listwise发挥潜力，超过pointwise和pairwise两者。因为当listwise遇到短列表（click number<grade number）时候，退化为pointwise，或者更差。
%而我们方法在这种不定长的文档个数时表现出鲁棒性，\textbf{因为query中点击文档的多少，只代表待判断的样例x的个数。}
(ii)when comparing these three datasets, we find that LLMs demonstrate the lowest performance on the KDD19 dataset. 
This is because KDD19 comes from a lab study where users complete complex search tasks\footnote{http://www.thuir.cn/KDD19-UserStudyDataset/}, making it challenging for both LLMs and humans to discern whether a document is useful for difficult tasks. 
(iii)when comparing these three methods, exhibits superior performance on datasets with longer average click lengths. The complexity of the KDD19 dataset results in longer average click lengths for queries, providing greater potential for listwise to outperform the other two approaches, which is consistent with the previous conclusions~\cite{zhang2024large}. However, when listwise is confronted with shorter click lists, its performance may even fall below that of pointwise. In contrast, our CLUE maintains robustness across varying document list lengths.
\vspace{-7pt}
\subsubsection{\textbf{Effects of Guideline}}
%我们做了消融实验来探究guideline的影响。就像前面提到的，我们的guidline 指导LLMs考虑这六个方面判断文档有用性。我将去掉这一指导的prompt变体叫做wo G，它在其他方面（例如打分策略）和我们的方法保持完全一致。
To verify the effectiveness of our guideline, we conduct an ablation study on the UUST and the KDD19 dataset. As mentioned earlier, our guideline guides LLMs to consider these six aspects when assessing document usefulness. The variant without this guideline is called w/o G. Other components are the same as those in our full model, CLUE.
% \begin{table}[h]
% \centering
% \caption{Performance of GPT on the three datasets with G or w/o G.}
% \vspace{-3mm}
% \label{tab:gpt2}
% \resizebox{\columnwidth}{!}{
% \begin{tabular}{lccccccc}
% \hline
% CLUE & \textbf{Precision} & \textbf{Recall} & \textbf{F1 Score} & \textbf{Pearson's \( r \)} & \textbf{Spearman's \( \rho \)} & \textbf{Cohen's \( \kappa \)} & \textbf{MAE} \\
% \hline
% \textbf{GPT-3.5}&  KDD19 & & & & & & \\
% \hline
%  w/o G & \textbf{0.3846} & 0.3751 & 0.3783 & 0.2911 & 0.2948 & 0.1448 & 1.0186 \\
%  w G & 0.3837 & \textbf{0.3805} & \textbf{0.3802} & \textbf{0.3159} & \textbf{0.3177} & \textbf{0.1511} & \textbf{1.0071} \\
% \hline
% \textbf{GPT-4} & & & & & & & \\
% \hline
%  w/o G & \textbf{0.3889} & 0.3630 & 0.3661 & 0.2946 & 0.2946 & 0.1430 & 1.0426 \\
%  w G& 0.3871& \textbf{0.3794}&\textbf{0.3668} &0.3457 & \textbf{0.3427}&  \textbf{0.1552}& \textbf{1.0489}\\
% \hline
% \textbf{GPT-3.5} &  UUST & & & & & & \\
% \hline
% w/o G & 0.4242 & 0.4196 & 0.4138 & 0.3590 & \textbf{0.3826 }& 0.1809 & 1.2056 \\
% w G & \textbf{0.4395} & \textbf{0.4252} & \textbf{0.4233} & \textbf{0.3751} & 0.3823 & \textbf{0.1911} & \textbf{1.1776} \\
% \hline
% \textbf{GPT-4 } & & & & & & & \\
% \hline
% w/o G & 0.4332 & 0.3902 & 0.4040 & 0.3775 & 0.3935 & 0.1672 & 1.2210 \\
% w G & \textbf{0.4419} & \textbf{0.4567 }& \textbf{0.4448} & \textbf{0.4275} & \textbf{0.4245} & \textbf{0.2273} & \textbf{1.1603 }\\
% \hline
% \end{tabular}}
% \end{table}
\begin{figure*}
  \centering
  \includegraphics[width=0.95\linewidth]{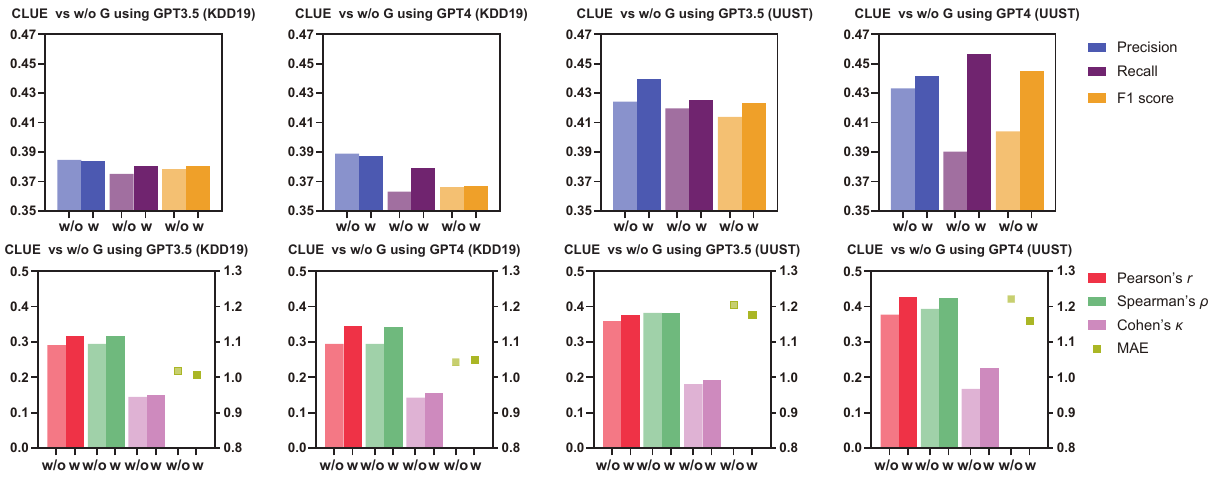}
  \vspace{-10pt}
  \caption{Comparison of the performance of the full model CLUE versus w/o G. The MAE references the right y-axis scale; other metrics reference the left.}
  \label{tab:ggg}
\end{figure*}
\vspace{-7pt}
\subsubsection*{\textbf{Results}:} The results are presented in  Figure ~\ref{tab:ggg}. Specifically, we find that (i) incorporating the guidelines obtained from the UUST dataset into prompts leads to notable improvements in performance on the UUST dataset. This highlights the potential of our guidelines to unlock the capabilities of LLMs in the usefulness judgment task. (ii) when applying the same guidelines to the KDD19 dataset, we also observed an overall enhancement in performance. This demonstrates the generalization capability of our guidelines and highlights the value of our new user study.

%%%%%%%%%%%%%%%%%%%%%%%%%%%%%%%%%%%%%%%%%%%%%%%%%%%%%%%%%%%%%%%%%%%1
% 在这里，我们需要写这个表格的 实验分析
% 我们report the agreement 不同设置下的prompt和 用户有些反馈之间的一致性 Across two dateset 

\begin{table}[h]
\centering
\caption{Usefulness judgments performance of Llama-3 before and after fine-tuning.}
\vspace{-10pt}
\label{tab:FTscore}
\resizebox{\columnwidth}{!}{
\begin{tabular}{p{1.5cm}cccccccc}
\hline
CLUE & Precision & Recall & F1-Score  & P-$r$ & S-$\rho$ & C-$\kappa$ & MAE\\ 
\hline
\textbf{SIGIR16} & & & & & & & \\
Llama-3 & 0.3164 & 0.3208 & 0.3128 & 0.2218 & 0.2202 & 0.0744 & 1.1726 \\
Llama-3 &\textbf{0.3537} & \textbf{0.3761} & \textbf{0.3576} & \textbf{0.2338} &\textbf{ 0.2446} & \textbf{0.1350} & \textbf{1.0907} \\ 
\hline
\multicolumn{2}{l}{\textbf{UUST}} & & & & & & \\
Llama-3 FT & 0.3948 &\textbf{ 0.3720 }& \textbf{0.3647} & 0.2218 & 0.2290 &\textbf{ 0.1180 }& 1.5608 \\
Llama-3 FT & \textbf{0.3958} & 0.3147 & 0.3412 & \textbf{0.3355} & \textbf{0.3231} & 0.0894 & \textbf{1.2867 }\\
\hline
\end{tabular}
}
\vspace{-12pt}
\end{table}
\vspace{-7pt}
\subsubsection{\textbf{Impact of Fine-tuning}:}
%{tab:prompt_str_train}
We compare two ways: (i) directly prompting Llama-3-8b-chinese-inst for judgment, and (ii) fine-tuning it with LoRA~\cite{dettmers2023qlora} using the training set detailed in Table~\ref{tab:shuju} and Section~\ref{sec:ft}, with parameters set to epoch=2 and max\_length=2048.
\vspace{-7pt}
\subsubsection*{\textbf{Results}:}
As shown in Table~\ref{tab:FTscore}, we find that: (i) fine-tuning improves performance, which proves benefits of using dedicated binary classifiers for each subproblem rather than a unified model.
(ii) neither Llama nor Llama-FT reach  GPT's capabilities. The performance gap between GPT and Llama is likely due to the limitation of maximum  context length. GPT-3.5 supports a context length of 16k, whereas our Llama is set to 2048, resulting in truncated document content.
\vspace{-5pt}
\subsection{RQ3: Satisfaction prediction: Using Usefulness-Based Framework}
\label{sec:expeRQ3}
%在满意度标注过程中，用户点击文档的质量（有用性/相关性）会对最终得分产生明显影响。通过基于有用性的框架，LLMs 的有用性判断能否为搜索查询满意度提供洞察力并改善满意度判断性能？(问题 3）。为了回答我们的 rq3.
 In this section, to validate the practical value of our usefulness judgments, we explore RQ3: Can we use LLM-based usefulness judgments from our method to enhance satisfaction prediction?%为了回答问题 3，我们在构建分类器和预测用户满意度时添加了基于有用性的特征。
To explore the question, we use the GBDT classifier with user behavior features as the baseline for satisfaction prediction, then, based on this, we add five click-sequence-based metrics by Mao et al. ~\cite{mao2016does}: $cCG$, $cDCG$, $cMAX$, $cCG/\#num$, $cDCG/\#num$.
%我们利用以下三个特征来预测满意度： (1) LLM 生成的相关性，(2) 第三方人工生成的相关性，以及 (3) LLM 生成的有用性。然后，我们通过计算这些特征的预测效果来分析它们对满意度的预测能力。我们使用分类器 GBDT进行预测分析。在本节中，我们将报告每个分类器的结果。
We compute them based on the following three labels: \textbf{(i)} LLM-generated relevance judgments \( R_{llm} \), \textbf{(ii)}  third-party relevance annotations \( R_{a} \), and \textbf{(iii)} LLM-generated usefulness judgments \( U_{llm} \). Then, we evaluate the predictive power of these features by comparing the performance of their classifiers. 
% \subsubsection{Common Experimental Setup in Section~\ref{sec:expeRQ3}}
%在之前的研究中，我们评估了LLM使用用户研究数据评估有用性的能力。我们还分析了不同组件对性能的影响，从而深入了解了如何有效地预测有用性。现在，我们探索LLM（特别是CLUE）的文档级判断如何增强查询级网络搜索评估，特别是在现实世界满意度预测方面。我们采用了完整的CLUE模型，特别是使用CULE的五选民版本，并且始终如一使用GPT-3.5来降低API费用。%在前两个RQ已经show了CLUE在有用性judge的优越性，在这个章节中，所有的$U_llm$指的都是LLM-driven judgments  from CLUE.
As we have already demonstrated CLUE’s suitability and superiority, we employ the full CLUE model to generate usefulness judgments, specifically using the five-voter version of CULE, and consistently using GPT-3.5 to lower API expenses. All $U_{llm}$ specifically refer to LLM-driven judgments from CLUE. 
\subsubsection{Experiment on the KDD19 dataset}
\begin{figure}[h]
\vspace{-16pt}
    \centering
    \begin{minipage}{0.45\columnwidth}
        \centering
        \includegraphics[width=\linewidth]{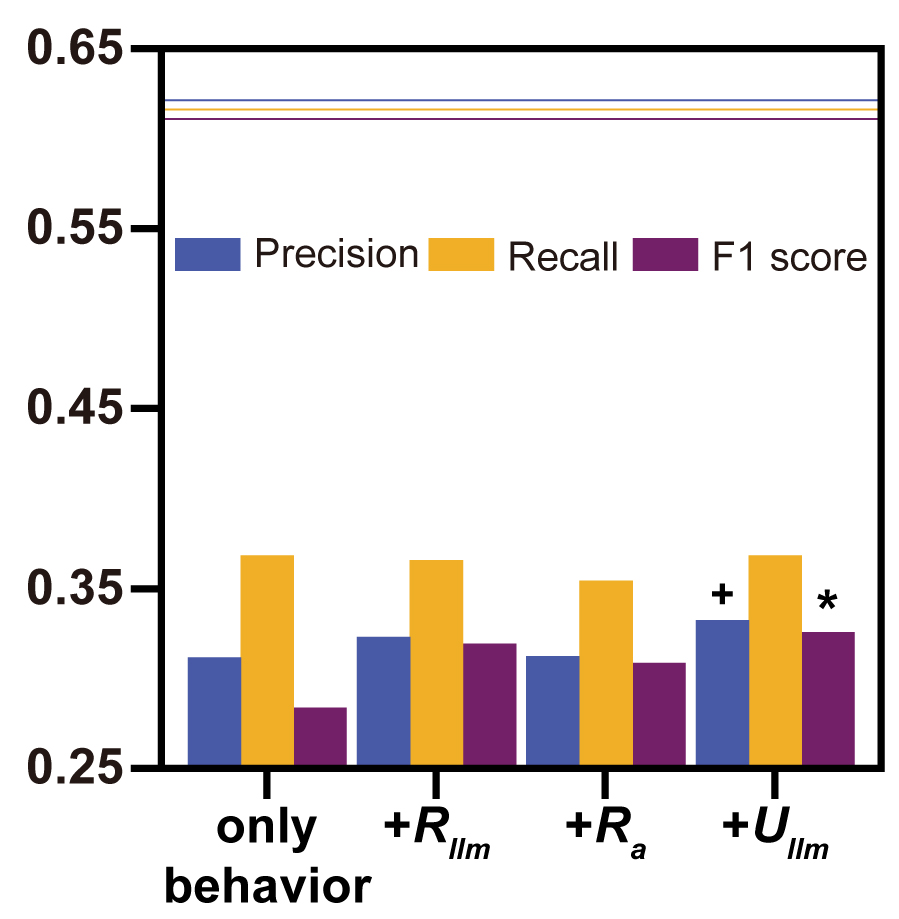} 
        \footnotesize{\begin{flushleft}
        "+" and "*" denote the method  significantly outperforms all baselines in paired t-test (+: p < .1, *: p < .05)  \end{flushleft}}
        \vspace{-10pt}
        \caption{Satisfaction prediction performance (KDD19)}
\label{fig:3KDD19}
    \end{minipage}%
    \hfill
    \begin{minipage}{0.5\columnwidth}
        \centering
        \vspace{-5pt} % 增加垂直间距，调整表格位置
        \resizebox{\linewidth}{!}{%
            \begin{tabular}{ccccc}
            \toprule
            \textbf{Metric}  & \textbf{$R_{a}$} & \textbf{$R_{llm}$} & \textbf{$U_{llm}$} \\
            \midrule
            P & - & 0.3579 & 0.3820 \\
            R & - & 0.3619 & 0.3751 \\
            F1 & - & 0.3591 & 0.3769 \\
            P-$r$  & 0.2579 & 0.2730  & 0.3087\\
            S-$\rho$ & 0.2692 & 0.2739  & 0.3127 \\
            C-$\kappa$ & - & 0.1166  & 0.1440\\
            MAE &  & 1.0459 & 1.0093 \\
            \bottomrule
            \end{tabular}
        }
        \footnotesize{\raggedright
        Note: $R_a$ cannot compute some metrics because it ranges from 0 to 3, whereas $U_u$ ranges from 1 to 4. }
        \captionof{table}{Performance of usefulness and relevance judgments using GPT-3.5 with t=0.7.}
        \label{tab:3KDD19}
    \end{minipage}
     \vspace{-14pt}
\end{figure}
%使用倒叙的手法，为什么不能提高这个的分数呢，是因为预测的相关性实在是和这个相距太远了。
%我们可以看到(i)skyline很高，使用真实标签Uu的模型取得了超级高的性能，但是加入相关性就不行。说明了1)使用有用性框架而非是相关性框架（Ra）才能准确评估用户满意度。2)第二点说明了用户质量变量对用户满意度的影响之大，这启示我们要想获得好的满意度预测性能，需要先得到$U_u$的好的替代指标。
%这里需要改
We build only behavior model use the same behavior features (mouse and keyboard-based measures and dwell time-based measures) extracted by Liu et al.~\cite{zhang2020investigating}, because these features are also commonly employed in previous IR evaluation studies~\cite{chen2017meta, jiang2015understanding}. %这里需要加入然后我们加入了三个特征。
The results shown in Figure~\ref{fig:3KDD19} suggest that the ranking of the consistency between the specified method and $U_u$ is \( U_{llm} >R_{llm} >R_a \). Similarly, the F1-score for satisfaction prediction also follow the
order \( U_{llm} >R_{llm} >R_a >   \text{only behavior} \). Moreover, $U_{\text{llm}}$ achieves significantly higher scores, and all three models outperform the model that uses only behavioral features.
This highlights: 
(i) the necessity of employing the usefulness framework ($+U_{llm}$) instead of the relevance-based approach ($+R_a$) for accurately assessing user satisfaction. As we can see, the skyline (colored lines in Figure~\ref{fig:3KDD19}), which uses the true labels $U_u$, achieves exceptionally high performance.
(ii) to achieve robust satisfaction prediction performance, it is crucial to first obtain reliable surrogate indicators for $U_u$. From Table ~\ref{tab:3KDD19} to Figure ~\ref{fig:3KDD19}, we can conclude that whoever accurately fits $U_u$ can achieve superior satisfaction prediction performance. 
(iii) our CLUE not only achieves the highest accuracy, but also benefits query-level satisfaction evaluation.
\vspace{-7pt}
 \subsubsection{Experiment on the SearchLog24Q3 dataset}
%可扩展的评估方式，我们对于这样一个可以扩展的前景非常光明的framework非常感兴趣，这样的一个能否用于实际的公司里，来改善搜索引擎的满意度预测问题中，对于文档内容建模不足的问题呢？我们使用来自最大的中文搜索平台百度的真实用户搜索日志和对应的满意度标注分数来验证我们的usefulness-based metric。在真实场景中，经过特征工程，用来预测满意度使用的日志数据特征主要包含用户在SERP页面上的点击、停留、翻页等行为，目前在用的特征有维度，目前在用的模型是gbdt。这就是满意度评估平台目前正在使用的真实场景。 used real search query between July and October 2024 for experimentation
%由于数据中没有人工生成的相关性标签$R_{a}$，这个数据列一直被预测的标签所占位in place of。因此经常使用百度开源模型ERNIE3.0-1.5B ~\cite{zhang2019ernie}预测的相关性标签。这是因为该模型与百度测试集上的人工标注数据相比，NDCG 达到了 0.87。
We validate our LLM-driven usefulness label in an anonymous search engine company. %然后，我们从这些日志中提取了用户的行为特征,仅包括query-level的行为特征和一些统计信息，如每个查询的停留时间、查询长度，用户是否翻页。我们总共提取了235个维度的特征。
The process and settings we employ are as follows:
(i) From logs, we extract 235 features, which are all behavioral features at the session level, query level, and aggregated measures at the document level. They are used to build `only behavior' model.%
%对应于日志数据中的每一个query，我们获取了用户当时搜索现场的每个文档自然语言内容，以及用户对该文档是否点击，点击的order是第几位，浏览的时长是多久等doc-level的特征，这将是我们输入到大模型的主要特征。
 (ii) For the usefulness judgments, we extract the same features as shown in Table ~\ref{tab:feature_for_llm}, and add them into the prompt as shown in Figure ~\ref{tab:prompt}.
%这里重复，都拿到这里来写
 (iii)%因为我们拿不到  改到这里了
 Due to the lack of human-generated relevance labels $R_{a}$, we use relevance labels (denoted as $R^{'}_{\text{a}}$) from the specialized relevance prediction model based on ERNIE 3.0-1.5B~\cite{zhang2019ernie}. 
 Because $R^{'}_{\text{a}}$ achieves an NDCG of 0.87 with human-generated relevance labels on the company's test set, we believe $R^{'}_{\text{a}}$ is a reliable alternative to $R_{\text{a}}$.
 %但是考虑越多的文档就会产生越大的API开销。我们在下边的

%对数据的清洗过程如下：我们去除了DQA结果，我们使用2024年7-10月的自然搜索的真实样本进行实验。In addition, 一些视频需求标签为3和图片需求为3的query，我们不能通过自然语言判断文档是否有用，因此我们在实验中过滤掉了这部分数据。我们还去除了色情query。最后，我们使用了9547条query真实数据进行实验。注意我们添加到大模型prompt中的特征都是在这235维特征中的，并没有添加其他新的特征进去。

\begin{table}[h]
\centering
\vspace{-5pt}
\caption{Satisfaction Classifier with different features on the real industrial search data. }
\label{tab:results}
\vspace{-10pt}
\resizebox{\columnwidth}{!}{
\begin{tabular}{p{2cm}p{1.2cm}p{1.2cm}p{1.2cm}p{1.2cm}p{1.2cm}p{1.2cm}}
\hline
 & Precision & Recall & F1-Score & MAE \\
\hline
only behavior & 0.5437 & 0.5625 & 0.5309 & 0.6667 \\ \hline
$+ R_{llm}$@3 & 0.5555 & 0.5874 & 0.5503 & 0.6014 \\
$+R_{llm}$@5 & 0.5770 & 0.5874 & 0.5542 & 0.6224 \\
$+R_{llm}$@10 & 0.5956 & 0.6084 & 0.5843 & 0.5804 \\ \hline
$+R^{'}_{\text{a}}$@3 & 0.5792 & 0.5972 & 0.5620 & 0.6181 \\
$+R^{'}_{\text{a}}$@5 & 0.5695 & 0.5972 & 0.5557 & 0.6111 \\
$+R^{'}_{\text{a}}$@10 & 0.5936 & 0.6111 & 0.5759 & 0.5694 \\ \hline
$+U_{llm}$@3 & 0.6046 & 0.6181 & 0.5817 & 0.5694 \\
$+U_{llm}$@5 & 0.5820 & 0.5903 & 0.5532 & 0.6250 \\
$+U_{llm}$@10 & \textbf{0.6301\textsuperscript{*}} & \textbf{0.6250\textsuperscript{*}} & \textbf{0.5915\textsuperscript{*}} & \textbf{0.5625} \\
\hline
\end{tabular}
}
\footnotesize{\begin{flushleft}
        "+" and "*" denote the model significantly outperforms all baselines in paired t-test (+: p < .1, *: p < .05)  \end{flushleft}}
\vspace{-25pt}
\end{table}
The results are shown in Table ~\ref{tab:results}.
%我们的加入有用性特征的分类器优于不加文档内容的纯用户行为的分类器，也持续优于了加入了基于大模型预测的相关性的特征的基线模型在@3和@5和@10截断上。这一结果表明，在对预测用户满意度的提升程度方面，大模型预测的有用性的预测性优于相关性的.More importantly, wat @3 and @10
Our classifier, which incorporates quality measures based on LLM-driven usefulness labels, achieves the best performance and consistently outperforms the baseline $+R_{llm}$ at @3, @5, and @10 cutoffs. It also consistently outperforms the baseline $+R^{'}_a$ at @3 and @10 cutoffs, and is comparable at the @5 cutoff. Overall, the degree of improvement led by quality labels from different sources is as follows: \( U_{llm} > R^{'}_a > R_{llm} > \text{only behavior} \). This conclusion validates both  the practical value of the usefulness-based framework and our LLM-driven judgments from CLUE.
\vspace{-9pt}
\section{Conclusion}
\label{sec:conc}
% 在这项研究中，我们探索了大语言模型（LLM）在信息检索场景中评估文档有用性的潜力。我们的研究表明，在这些任务中，结合有用性判断可以提高满意度预测模型的性能，LLM 生成的有用性已经能够取代第三方相关性注释。这为通过整合 LLM 生成的有用性标签来改进搜索技术提供了一个很有前景的方向。

% 作为对传统搜索引擎评估中 LLM 实用性判断的初步研究，我们只尝试了相对简单和幼稚的提示来判断有用性，在提高判断性能方面仍有一些策略可以探索，例如思维链提示（Chain-of-Thought Prompting）。此外，我们的分析仅限于一组用户搜索日志，这取决于用户与检索系统的现有交互。未来的工作应侧重于使用模拟不同用户配置文件的方法来生成搜索日志。这种从搜索引擎到用户满意度的全自动框架极具吸引力。

% 我们希望我们的工作能为 LLM 的未来发展和实用性判断的部署提供有价值的评估框架和有洞察力的发现。我们期待进一步的研究能为 LLM 的实用性评估领域做出重大贡献，最终实现更有效的、以用户为中心的信息检索系统评估。

% In this study, we explored the potential of Large Language Models (LLMs) to assess document usefulness in information retrieval scenarios.  我们设计了prompt包含q,d,i去instruct大模型判断有用性，并且为这个分级分类问题设计了三种不同输入format 的prompt。
% Our research demonstrates that by incorporating contextual information, well-instructed LLMs  can step into the
% user’s shoes and judge user-perceived usefulness. Moreover, 我们做了新的用户实验in order to搜集了用户对usefulness的思考来更好的指导大模型判断有用性。评测结果表明我们从用户思考中总结的guideline提升了判断的准确性同时具有泛化性。我们还使用了微调来获得更精准的有用性打分和思考。最后我们测试了有用性标签的实际价值in real-world，我们用它提升了百度公司满意度预测的性能。
In this study, we explore the potential of utilizing Large Language Models (LLMs) to judge document usefulness in information retrieval scenarios. We meticulously design prompts specifically for usefulness judgment and propose a new method CLUE tailored for this task, which incorporate multiple components and surpass third-party annotations in terms of accuracy.  Secondly, we conduct a new user study to collect user thoughts on usefulness, in order to better guide the LLM’s judgment. Evaluation results show that the guideline improve the accuracy. We also fine-tune an open-source LLM to enhance performance. Finally, with both user study data and a real-world search log, we test the practical value of these usefulness labels in RQ3.

Our exploration is still in its early stages, there are still strategies that can be explored, such as Chain-of-Thought prompting. Additionally, our analysis is limited to a set of user search logs, which depend on existing user interaction with the retrieval system.

We hope our work contributes to the user-centric evaluation of IR systems. For future work, we can implement this usefulness-based framework in practical applications. For instance, we can obtain large-scale usefulness judgments through this approach, and then apply the satisfaction prediction described in section ~\ref{sec:expeRQ3} to conduct query-level evaluation. Additionally, We can combine user simulation methodologies to generate search logs and apply the approach in section ~\ref{sec:expeRQ3} to automatically evaluate user satisfaction before real users interact with it. This fully automated framework—spanning from the search engine to user satisfaction—holds significant potential. Last but not least, we hope that the initial insights from our preliminary efforts, combined with the new user dataset, will contribute the future work on LLM-based usefulness judgment.
\vspace{-10pt}
\appendix
\section{Dataset Collection}
\subsection{User Study}
\label{tab:Dataset Collection}
%为了更好地了解用户的真实想法，我们开展了一项用户研究。为了直接观察用户的思考过程，当用户报告有用性得分时，我们要求他们通过大声思考（think aloud/cite[]）来提供自己的想法。Collecting and analyzing think-aloud data has been used to build models of cognitive processes during a problem solving task[35].  While doing the task, we asked participants to say out loud what goes through their mind by stating directly what they think.这种实时思考有助于我们深入了解用户的决策过程，更好地理解影响用户判断的因素。
To better understand the real thoughts of users, we conduct a user study where participants are asked to think aloud while reporting usefulness scores, following the think-aloud protocol ~\cite{van1994think}. Collecting and analyzing think-aloud data is a method commonly used to develop models of cognitive processes during problem-solving tasks ~\cite{ghenai2020think, rethinking2018}. By this approach, real-time reflection helps us gain deeper insights into their decision-making process and better understand the factors influencing their judgments.

%在每个任务开始前，我们需要用户记住并复述任务描述，尤其是任务的信息需求。完成这些初步步骤后，参与者开始进行主要的搜索任务。
Before starting each task, we require users to remember and restate the task description, especially the information needs of the task. 
%在整个搜索过程中，参与者需要在大声思考的同时完成任务。同时我们进行录音和录像，插件记录用户的搜索行为。
%每完成一个任务后，参与者会被要求标注他们点击的网页的有用性，并提供最终的query-level的满意度评分。
They are required to complete these search tasks while thinking out aloud. At the same time, we begin recording both video and audio, and a browser plugin begins logging users' search behaviors. After completing each task, the participant is asked to annotate the usefulness of the documents they click on and provide final query-level satisfaction.

%在参与者完成所有任务后，我们整理了他们在关键时间节点的录音，包括输入查询时、点击链接时、有用性标注时和满意度标注时，并通过录屏来核实记录的语音时刻是否准确。
Once all tasks are completed, we organize and manually transcribe the audio recordings from key time points—when participants annotate usefulness and provide satisfaction ratings—and cross-check the timestamps with the corresponding screen recordings.
%最后，我们对数据集进行了整理。为了避免实验过程中检索出的文档链接失效，我们在实验结束后立即进行了网页截图和OCR内容保存。
After the experiment, to get the full document content, we immediately take screenshots of the web pages and perform OCR to save the content. Finally, we name this dataset: User Usefulness and Satisfaction Thought (UUST).
\vspace{-9pt}
\subsection{Search Task Design and User Recruitment}
%10个任务包含简单，中等和complex的任务被抽取出来从类别，包括医疗咨询、作业辅导、专家知识等。图3展示了一些选出来的任务。
We design 10 tasks covering numerous categories like medical inquiries, homework assistance, and computer knowledge. Table~\ref{tab:task_details} provides an example of these tasks. satisfaction feedback.
\begin{table}[h]
\vspace{-10pt}
\caption{Task Details}
\vspace{-10pt}
\label{tab:task_details}
\centering
\resizebox{\columnwidth}{!}{
\begin{tabular}{p{5cm}|p{5cm}}
\hline
\textbf{Background} & \textbf{Goal} \\
\hline
You are very interested in environmental protection and sustainable development. You want to learn about some practical sustainable lifestyle practices and recommendations. & Based on the search results, please provide three sustainable lifestyle practices and recommendations and explain their positive impact on the environment. \\
\hline
\end{tabular}}
\vspace{-12pt}
\end{table}
%我们通过社交网络招募了34名参与者，最终留下了31位参与者的数据。这31位都是学校的在读学生（本科硕士和博士），参与者中女性有18人，男性有13人。
We recruit 34 participants through social networks, and ultimately, data from 31 participants are retained. All 31 participants are current students (undergraduate, master's, and doctoral levels). Among the participants, 18 are female, and 13 are male. 
%我们重新设计了 Pogacar 等人[25]的研究界，它是a traditional style of web search engine.
We recreate the Mao et al. ~\cite{mao2016does} interface , a traditional web search engine style, for our study. 
%用户可以像使用商业搜索引擎一样使用这个实验性搜索引擎。 并且，每次用户提交查询时，SERP 页面都会实时显示从搜狗商业搜索引擎检索到的十个搜索结果。我们使用浏览器插件记录用户的搜索互动，包括所有点击和阅读时长。
The participant can use this experimental search engine in the same way they normally do when using commercial search engines, and each time a user submits a query, the SERP page displays ten search results retrieved from a commercial search engine in real time. We use a browser plugin to log users' search interactions, including all clicks and movements. 
\vspace{-9pt}
\section{Implements Detail of GBDT for Usefulness}
\label{tab:task_details}
% 4.14这里添加：由于我们希望利用在线环境中可获得的行为变量来预测有用性判断，因此我们去除了“前一页面的最大有用性”、“任务难度”、“任务领域知识”以及“任务兴趣”等特征变量。我们选择上下文特征变量如图所示：max_usefulness_of_previous_page,task_difficulty,task_domain_knowledge,task_interest
% 另外，我们的行为变量不包含avg_eye_speed，fixation_num 等需要使用眼动仪的，不包含mouse_movement_num在内容页上的变量。少了这几个和最后有用性强相关的变量，因此我们的性能不如他们报告的。我们在不同数据集上复现了他们的性能，使用了相似的设备和相同的插件，避免了领域偏移的问题。然而，使用小数据集建立机器学习模型时，其泛化能力面临挑战。
% 为了维护比较时的公平性，我们在复现他们的方法时，除了我们方法中使用的特征（如表2所示），我们还加入了他们提出的特征（在sigir17论文的表1表2表3）。
For fairness comparison, features used in our CLUE (Table~\ref{tab:feature_for_llm}) and in their work~\cite{mao2017understanding} (in their Tables 1, 2 and 3, but eye-tracking and mouse movement/scrolling data could not be obtained for us, we only can obtain subset as shown in Table~\ref{tab:GBDTfeatures} ) are both used to establish the  GBDT classifier.
% 但我们拿不到眼动数据，which在除了实验室以外的环境中很难拿到。除此之外，细粒度的鼠标移动、滚动数据我们拿不到，scroll_num，mouse_movement_num，这种变量提取不了。
Moreover, we performed grid-search over the n\_estimators, learning\_rate, max\_depth, and subsample, and report the best results.% over the n\_estimators, learning\_rate, max\_depth, and subsample, and reported the best results.
\begin{table}[ht]
    \centering
    \caption{Extracted Features for GBDT Model.}
    \vspace{-10pt}
    \label{tab:GBDTfeatures}
    \resizebox{\columnwidth}{!}{
    \begin{tabular}{@{}lll@{}}
        \toprule
        \textbf{Content Factors} & \textbf{Context Factors} & \textbf{Behavior Factors} \\ \midrule
        content\_bm25\_with\_query & num\_previous\_doc & page\_dwell\_time \\
        content\_cossim\_with\_query & num\_previous\_doc\_in\_query & query\_length \\
        content\_cossim\_with\_task\_description & num\_previous\_query & query\_time \\
        & progress\_time\_in\_session & session\_time \\ \bottomrule
    \end{tabular}}
\end{table}
\section{GenAI Usage Disclosure}
Our paper does not include text generated by a Large Language Model. We only used a Large Language Model for automatic grammar checks and word corrections.

\bibliographystyle{ACM-Reference-Format}
\bibliography{sample-base}

\end{document}